\documentclass[%
reprint,
superscriptaddress,
amsmath,amssymb,
aps,
prb,
floatfix
]{revtex4-2}

\usepackage[pdftex]{graphicx}
\usepackage{bm}
\usepackage{amsmath}
\usepackage{textcomp}
\usepackage{xr-hyper}
\usepackage{hyperref}
\usepackage[all]{hypcap}
\usepackage{xcolor}

\begin{document}
\title{Determination of the current-phase relation of an InAs 2DEG Josephson junction with a microwave resonator}
\makeatletter
\let\newtitle\@title
\let\newauthor\@author
\let\newdate\@date
\makeatother

\author{Zoltán Scherübl}
\thanks{These authors have contributed equally to this work.}
\affiliation{Department of Physics, Institute of Physics, Budapest University of Technology and Economics, M\H uegyetem rkp.\ 3., H-1111 Budapest, Hungary}
\affiliation{MTA-BME Superconducting Nanoelectronics Momentum Research Group, M\H uegyetem rkp.\ 3., H-1111 Budapest, Hungary}

\author{Máté Sütő}
\thanks{These authors have contributed equally to this work.}
\affiliation{Department of Physics, Institute of Physics, Budapest University of Technology and Economics, M\H uegyetem rkp.\ 3., H-1111 Budapest, Hungary}
\affiliation{MTA-BME Superconducting Nanoelectronics Momentum Research Group, M\H uegyetem rkp.\ 3., H-1111 Budapest, Hungary}

\author{Dávid Kóti}
\affiliation{Department of Physics, Institute of Physics, Budapest University of Technology and Economics, M\H uegyetem rkp.\ 3., H-1111 Budapest, Hungary}
\affiliation{MTA-BME Superconducting Nanoelectronics Momentum Research Group, M\H uegyetem rkp.\ 3., H-1111 Budapest, Hungary}

\author{Endre T\'ov\'ari}
\affiliation{Department of Physics, Institute of Physics, Budapest University of Technology and Economics, M\H uegyetem rkp.\ 3., H-1111 Budapest, Hungary}
\affiliation{MTA-BME Correlated van der Waals Structures Momentum Research Group, M\H uegyetem rkp.\ 3., H-1111 Budapest, Hungary}

\author{Csaba Horváth}
\affiliation{Department of Physics, Institute of Physics, Budapest University of Technology and Economics, M\H uegyetem rkp.\ 3., H-1111 Budapest, Hungary}
\affiliation{MTA-BME Superconducting Nanoelectronics Momentum Research Group, M\H uegyetem rkp.\ 3., H-1111 Budapest, Hungary}

\author{Tamás Kalmár}
\affiliation{Department of Physics, Institute of Physics, Budapest University of Technology and Economics, M\H uegyetem rkp.\ 3., H-1111 Budapest, Hungary}
\affiliation{MTA-BME Superconducting Nanoelectronics Momentum Research Group, M\H uegyetem rkp.\ 3., H-1111 Budapest, Hungary}

\author{Bence Vasas}
\affiliation{Department of Physics, Institute of Physics, Budapest University of Technology and Economics, M\H uegyetem rkp.\ 3., H-1111 Budapest, Hungary}
\affiliation{MTA-BME Superconducting Nanoelectronics Momentum Research Group, M\H uegyetem rkp.\ 3., H-1111 Budapest, Hungary}

\author{Martin Berke}
\affiliation{Department of Physics, Institute of Physics, Budapest University of Technology and Economics, M\H uegyetem rkp.\ 3., H-1111 Budapest, Hungary}
\affiliation{MTA-BME Superconducting Nanoelectronics Momentum Research Group, M\H uegyetem rkp.\ 3., H-1111 Budapest, Hungary}

\author{Magdhi Kirti}
\affiliation{CNR - Istituto Officina dei Materiali (IOM), Area Science Park Basovizza, Trieste, 34149, Italy}

\author{Giorgio Biasiol}
\affiliation{CNR - Istituto Officina dei Materiali (IOM), Area Science Park Basovizza, Trieste, 34149, Italy}

\author{Szabolcs Csonka}
\email{csonka.szabolcs@ttk.bme.hu}
\affiliation{Department of Physics, Institute of Physics, Budapest University of Technology and Economics, M\H uegyetem rkp.\ 3., H-1111 Budapest, Hungary}
\affiliation{MTA-BME Superconducting Nanoelectronics Momentum Research Group, M\H uegyetem rkp.\ 3., H-1111 Budapest, Hungary}

\author{P\'eter Makk}
\email{makk.peter@ttk.bme.hu}
\affiliation{Department of Physics, Institute of Physics, Budapest University of Technology and Economics, M\H uegyetem rkp.\ 3., H-1111 Budapest, Hungary}
\affiliation{MTA-BME Correlated van der Waals Structures Momentum Research Group, M\H uegyetem rkp.\ 3., H-1111 Budapest, Hungary}

\author{Gerg\H{o} F\"ul\"op}
\affiliation{Department of Physics, Institute of Physics, Budapest University of Technology and Economics, M\H uegyetem rkp.\ 3., H-1111 Budapest, Hungary}
\affiliation{MTA-BME Superconducting Nanoelectronics Momentum Research Group, M\H uegyetem rkp.\ 3., H-1111 Budapest, Hungary}

\date{\today}
\begin{abstract}
Semiconductor-superconductor hybrid nanocircuits are of high interest due to their potential applications in quantum computing. Semiconductors with a strong spin-orbit coupling and large $g$-factor are particularly attractive since they are the basic building blocks of novel qubit architectures. However, for the engineering of these complex circuits, the building blocks must be characterized in detail. 
We have investigated a Josephson junction where the weak link is a two-dimensional electron gas (2DEG) hosted in an InAs/InGaAs heterostructure grown on a GaAs substrate.
We employed the in-situ epitaxially grown Al layer as superconducting contacts to form an rf SQUID, and also to create a microwave resonator for sensing the Josephson inductance.
We determined the gate-dependent current-phase relation, and observed supercurrent interference in out-of-plane magnetic fields. With the application of an in-plane magnetic field, we induced asymmetry in the interference pattern, which was found to be anisotropic in the device plane.

\end{abstract}

\maketitle

\section{\label{sec:intro}Introduction}

Superconducting (SC) electronic circuits and qubit architectures have gone through an immense development over the last years\,\cite{Kjaergaard2020}.
By now architectures with qubit numbers in the range of 1000 have been engineered, however, the limited lifetime of the qubits sets a serious bottleneck. To resolve this, novel devices are being developed, including superconductor-semiconductor hybrids. Here the remarkable properties of the superconducting circuits are combined with the gate tunability of the semiconductors, leading to the implementation of gatemons\,\cite{PhysRevLett.115.127001, PhysRevResearch.6.023094, PhysRevLett.115.127002,  Casparis2018,Zheng2023CoherentCO, Sagi2024}, Andreev qubits\,\cite{Hays2020, Hays2021, Pita2023, Pita2024}, or gatemonium\,\cite{Strickland2024}. 
More importantly, further proposals suggest that the combination of spin-orbit coupling (SOC) and SC properties can lead to systems where the information is protected by the topological nature of the qubit, enabling fault-tolerant quantum computing\,\cite{Mourik2012, Sarma2015, Deng2016, Aguado2017, Fornieri2019, Dartiailh2021, Amundsen2024}.
Despite intense work on semiconductor nanowires and two-dimensional electron gases (2DEG) to realise  Majorana-based qubits, this has not been undoubtedly achieved.
Along similar lines, recent works using a bottom-up approach to realise Kitaev chains based on quantum dot-SC arrays have reached important milestones\,\cite{Sau2012, Leijnse2012, Zatelli2023, Dvir2023, Sebastiaan2024}. 

The central building blocks of these architectures are SC-N-SC Josephson junctions (JJ), where in the normal (N) region Andreev bound states (ABS) form, which carry the supercurrent.
These states have been investigated in a wide range of systems using tunnel probes\,\cite{Buitelaar2002, Pillet2010, Lee2013, Jellinggaard2016, Bretheau2017, Wang2018a, Prada2020, vanDriel2023}, microwave spectroscopy\,\cite{Bretheau2013, Woerkom2017, Dassonneville2018, Tosi2019,  Chidambaram2022, Hinderling2023} and SC current-phase measurements\,\cite{Szombati2016, Nanda2017, Spanton2017open, PhysRevLett.99.127005, Indolese2020, Haller2022, Haxell2023,  Haxell2023}.
Since SOC is required for the realisation of Majorana states, the N region is made from semiconductors with large SOC.

An InAs 2DEG offers a versatile and scalable platform with a sizable SOC\,\cite{Strickland2022, Elfeky2023, Phan2022, Phan2023, Serafim2024}. 
When the 2DEG is designed to form close to the surface, it can be coupled strongly to SC electrodes. However, to realise topological systems and use them in fast quantum information processing schemes, the understanding of ABS and their coupling to radio-frequency (rf) circuitry is needed.
The role of higher harmonics and spin-orbit properties in the current-phase relation (CPR) are important for conventional SC-based or protected qubits as well. For example, it has recently been demonstrated that the spectrum of tunnel-junction-based transmons is detectibly modified by the higher harmonic content of the CPR, which is usually assumed to be perfectly sinusoidal in tunnel junctions\,\cite{Willsch2024}. 

We have recently shown that an InAs 2DEG coupled to SC electrodes can be grown on GaAs substrates with high mobility and good proximity effects\,\cite{Suto2022}.
Here we investigate ABS in JJs formed on this platform using high-frequency techniques. We take advantage of the in-situ grown aluminium to form SC resonators and perform CPR measurements on inductively coupled Josephson junctions for different doping levels. We find non-sinusoidal CPR at larger doping and more sinusoidal behaviour close to pinchoff. Using the SC resonator, we are also able to map out the interference pattern of the junction in an out-of-plane magnetic field showing a Fraunhofer-like pattern. Finally, we show the decrease of the critical current $I_c$ and the appearance of a marked anisotropy in an in-plane magnetic field.

\section{\label{sec:sample}Sample fabrication}

The samples were realized in an InAs/InGaAs semiconducting heterostructure 
grown with molecular beam epitaxy on a GaAs substrate\,\cite{Benali2022, Suto2022}. The semiconducting wafer hosts a 2DEG buried 10\, nm below the surface in an InAs quantum well, which is contacted by a 50-nm-thick epitaxial Al layer.
We employ this Al layer to define a SQUID loop and a microwave resonator for the dispersive sensing of the Josephson inductance. We used electron beam lithography (EBL) in combination with wet chemical etching for the selective removal of the Al layer and mesa formation in the semiconductor. In the next lithography step we defined the Josephson junction weak link by the local etching of the Al in the SQUID loop.
For gating, we deposited a 50-nm-thick Al$_2$O$_3$ dielectric layer with atomic layer deposition (ALD), covering the entire sample, and used EBL to create the Ti/Au gate electrode on top of the junction. For more details of the sample fabrication, see the Supplementary Material (SM).    

\begin{figure}[htb]
\includegraphics{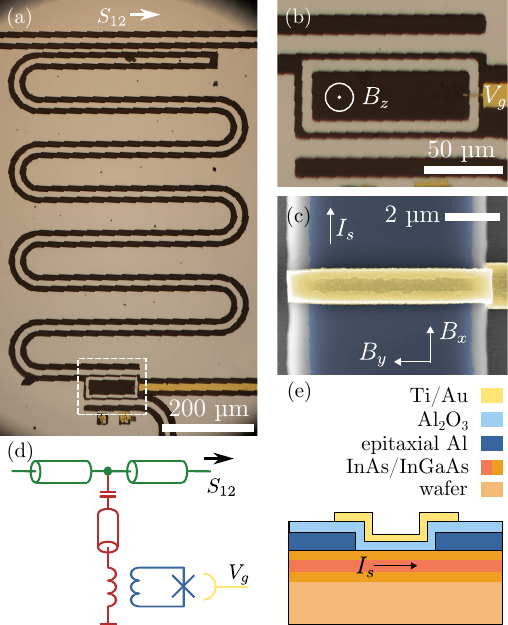}
\caption{\label{fig:sample} Sample geometry and electrical circuit model. (a) Optical image of the $\lambda/4$-resonator, coupled to the feedline at the open end (top), and coupled to the rf SQUID loop at the shorted end (bottom).
(b) Zoom-in to the rf SQUID area. The Josephson junction is embedded in the epitaxial Al loop, with a gate allowing the tuning of the charge carrier density in the junction. Near the bottom of the loop, an on-chip flux line is present (not used).
(c) False-colored SEM image of the Josephson junction (device A): epitaxial Al is dark blue, the Ti/Au gate electrode is yellow.
(d) Circuit model of the coupled resonator-SQUID system. The feedline is depicted in green, the resonator in red, the rf SQUID in blue, the gate in yellow.
(e) Cross section of the Josephson junction, with a simplified representation of the semiconducting heterostructure (not to scale). The epitaxial Al layer (dark blue) forms ohmic contacts to the 2DEG (red) in the semiconducting InAs/InGaAs heterostructure.}
\end{figure}

We show the resulting sample geometry in Fig.\,\ref{fig:sample}. In this paper we discuss measurement data from two nominally identical devices; the images in Fig.\,\ref{fig:sample}(a-c) were taken of device A.
Panel (a) is an optical image, showing at the top a coplanar waveguide (CPW) feedline on which the microwave transmission $S_{12}$ is measured. A meandering quarter-wave CPW resonator is coupled capacitively to the feedline at the open end, and inductively to an rf SQUID at the shorted end. The SQUID loop acts as the load of the resonator, and enables sensing the JJ impedance through the microwave readout of the resonator.
Fig.\,\ref{fig:sample}(b) shows a zoom-in to the SQUID area. The JJ is embedded in the right segment of the loop, and is covered by the gate. The inner dimensions $l_{x,y}$ of the loop are 30.5\,\textmu m $\times$ 101\,\textmu m. The phase of the JJ can be tuned by threading magnetic flux in the loop with an out-of-plane magnetic field $B_z$. A scanning electron microscope (SEM) image of the Josephson junction is shown in Fig.\,\ref{fig:sample}(c). In device A the junction had a length $L=450\,$nm and width $W=6.2\,$\textmu m, while for device B, $L=400\,$nm and $W=8.5\,$\textmu m. Arrows indicate the choice of the coordinate system: the $x-y$ plane lies in the sample plane, with $x$ being parallel to the supercurrent $I_s$, and the [110] crystallographic direction. 

Fig.\,\ref{fig:sample}(d) shows the equivalent electrical circuit of a single device. We measure the microwave scattering parameter $S_{12}$ on the feedline (green) to which the hanger-mode resonator (red) is coupled. The resonator is coupled inductively to the SQUID loop (blue), in which the JJ is tuned by applying the gate voltage $V_g$ on the gate electrode (yellow).

In Fig.\,\ref{fig:sample}(e) we illustrate the simplified cross section of the JJ. The exact heterostructure composition of the semiconducting wafer is published in Refs. \citenum{Suto2022} and \citenum{Benali2022}. The charge transport takes place in the InAs layer, which is proximitized by the epitaxial Al layer on the surface.

\section{\label{sec:results}Experimental results}
\subsection{\label{sec:cpr}Determination of the current-phase relation}

All measurements were carried out in a dilution refrigerator at $T\approx100\,$mK. Above the sample chip we mounted an aluminium plate as a shield against external magnetic noise to reduce flux fluctuations in the SQUID loop. This provided only partial shielding and still allowed us to apply a magnetic field on the sample with the superconducting magnet of the refrigerator. We measured the microwave transmission on the feedline with a vector network analyzer, and applied the gate voltage $V_g$ through a filtered DC line with a DC voltage source. The details of the measurement setup can be found in the SM.

\begin{figure}[htb]
\includegraphics{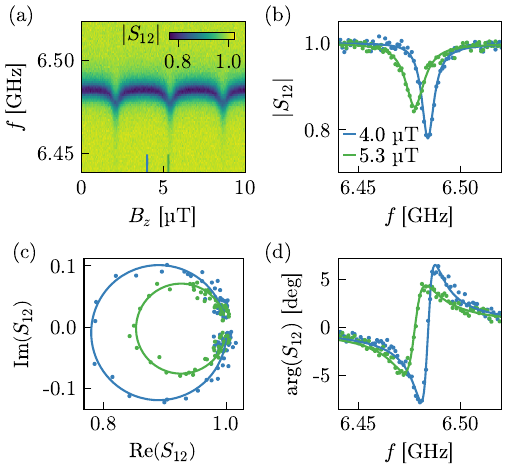}
\caption{\label{fig2} Microwave transmission measurements at $V_g=-7\,$V (device A). (a)  Transmission magnitude $|S_{12}|$ as a function of out-of-plane magnetic field $B_z$ and frequency $f$. (b) Magnitude and (d) phase as a function of frequency near resonance at $B_z=4\,$\textmu T (blue) and $5.3\,$\textmu T (green), as marked by colored lines in panel (a). Data points are plotted as dots, fitted curves as solid lines. (c) The same two resonances shown in the IQ plane.}
\end{figure}

In Fig.\,\ref{fig2} we show the normalized microwave transmission $S_{12}$ measured on the feedline. The resonance of the quarter-wave resonator manifests in a local minimum in $|S_{12}|$ as a function of frequency $f$, which shifts periodically in the magnetic field $B_z$ as presented in panel (a). The shift of the resonance frequency originates from the flux-tuning of the rf SQUID, i.\,e. tuning of the Josephson inductance of the junction. From the loop geometry, the expected periodicity of the shift is $\Delta B_z = \Phi_0/l_x l_y = 0.67\,$\textmu T. Experimentally, we observe a period of $\Delta B_z = 3.26\,$\textmu T. The almost 5-fold difference originates from the employed Al shield. In reference measurements without magnetic shielding we confirmed that the measured periodicity is consistent with the theoretical expectation based on the nominal loop area. 

For the quantitative evaluation, we fit the resonances of the notch-coupled $\lambda/4$-resonator with the complex transmission function\,\cite{khalil2012, Probst2015,Wang2021} 
\begin{equation}
\label{eq:resonance}
S_{12} (f) = 1 - \frac{Q e^{i\vartheta}}{Q_c (1 + 2i Q (f-f_0)/f_0)},
\end{equation}
where $Q=1/(\cos{(\vartheta)}/Q_c + 1/Q_i)$. The fit parameters are the resonance frequency $f_0$, the internal quality factor $Q_i$, the coupling quality factor $Q_c$, and the resonance asymmetry parameter $\vartheta$. In this partitioning, the loss in both the resonator and the rf SQUID is incorporated in $Q_i$.

Two fitted resonance curves, selected at $B_z=4\,$\textmu T (blue) and $5.3\,$\textmu T (green) are illustrated in Fig.\,\ref{fig2}(b) and (d), by plotting the magnitude and phase of the complex transmission as a function of $f$. Fig.\,\ref{fig2}(c) shows the data points and the fit in the IQ plane. The resonance frequency and the depth both change, indicating that the resonator load has a dispersive and a dissipative response as well.

\begin{figure}[htb]
\includegraphics{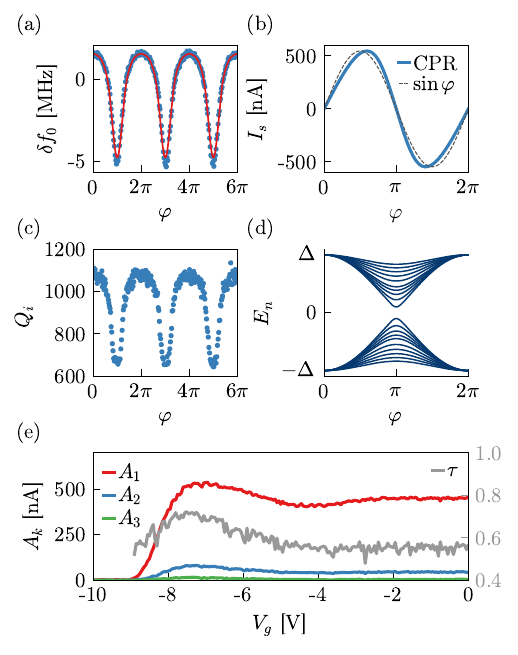}
\caption{\label{fig3} Determination of the current-phase relation and harmonic coefficients (device A). (a) Resonance shift $\delta f_0$ as a function of junction phase $\varphi$ at $V_g=-7\,$V. (b) The reconstructed current-phase relation (solid blue), with a sinusoidal function as a reference (dashed gray). (c) Internal quality factor $Q_i$ as a function of $\varphi$ at $V_g=-7\,$V. (d) Qualitative illustration of the Andreev bound state spectrum of the JJ. (e) Harmonic coefficients $A_k$ (scale on the left) and effective transmission $\tau$ (scale on the right) as a function of gate voltage $V_g$.}
\end{figure}

We plot the fit results $\delta f_0 = f_0 - f_{\rm{bare}}$ and $Q_i$ as a function of the phase $\varphi$ of the junction in Fig.\,\ref{fig3}(a,c), respectively. The calculation of the offset $f_{\rm{bare}} \approx 6.482\,$GHz and $\varphi(\Phi_{\rm{ext}})$ where $\Phi_{\rm{ext}}$ is the applied flux is part of the self-consistent approach discussed below. We use the frequency shift $\delta f_0$ to determine the CPR of the junction following the method developed in Ref.\,\citenum{Haller2022}, which takes into account the magnetic flux created by the supercurrent circulating in the SQUID loop. In this framework, the coplanar waveguide resonator is modeled with an equivalent LC-circuit with inductance $L_p$. The resonator is coupled inductively to the SQUID with mutual inductance $M$. The SQUID acts as the load of the resonator, where the superconducting loop itself has a self-inductance of $L_{\rm{loop}}$, and embeds the junction impedance $Z_J$. The change in the inductive part $L_J$ of this impedance results in the relation 
\begin{equation}
\label{eq:fres}
\delta{}f_0 (\varphi) \approx\frac{8}{\pi^2}\frac{M^2}{L_p\left(L_J (\varphi) +L_{\rm{loop}}\right)}f_{\rm{bare}},
\end{equation}
where $f_{\rm{bare}}$ is the resonant frequency of the unloaded resonator. The junction inductance can be expressed as
\begin{equation}
\label{eq:JJ_inductance}
L_J(\varphi)^{-1}=\frac{2\pi}{\Phi_0}\frac{\partial I_s(\varphi)}{\partial\varphi},
\end{equation}
where $I_s(\varphi)$ is the CPR and $\Phi_0=h/2e$ is the magnetic flux quantum. We express the CPR as a Fourier series with only sinusoidal terms\,\cite{Strambini2020}, 
\begin{equation}
\label{eq:IsFourier}
I_s(\varphi)=\sum_{k=1}^{k_{\rm{max}}}(-1)^{k-1}A_{k}\sin(k\varphi).
\end{equation}
Theoretically, in the case of a tunnel junction with low transmission, only the $k=1$ term is significant and the CPR is sinusoidal. For higher transmission, the CPR is non-sinusoidal, which is reflected in the non-zero higher-order coefficients, which typically diminish with increasing $k$. In our model $k_\mathrm{max}=6$, which proves to be sufficient since $A_k$ is negligible for $k>3$ as shown later. We include the alternating sign in the series expression so that for a forward-skewed CPR the coefficients $A_k$ are positive.
The analytical form of the CPR allows us to write $\delta f_0(\varphi)$ in a closed form for the fitting to the experimental data by substituting equations (\ref{eq:IsFourier}) and (\ref{eq:JJ_inductance}) into Eq. (\ref{eq:fres}).

The flux created by the supercurrent circulating in the SQUID loop modifies the phase as
\begin{equation}
\label{eq:screening}
\varphi=\varphi_{\rm{ext}}-\frac{2\pi}{\Phi_0}L_{\rm{loop}}I_s(\varphi),
\end{equation}
where $\varphi_{\rm{ext}}=2\pi\Phi_{\rm{ext}}/\Phi_0$. This screening effect results in a non-linear transformation of the junction phase and, for the precise determination of the CPR shape, must be taken into account. Thus, we combine the curve fitting with fixed-point iteration using Eq. (\ref{eq:screening}) to obtain a self-consistent solution.

The resulting fit of $\delta f_0(\varphi)$ is shown in Fig.\,\ref{fig3}(a) in a red line. Using the gained fit parameters, we plot the CPR in Fig.\,\ref{fig3}(b). The CPR is slightly forward skewed compared to a sinusoidal function (dashed), with a critical current of $I_c\approx550\,$nA.

The internal quality factor $Q_i$ varies in the range of 650--1100, depending on the junction phase, as shown in Fig.\,\ref{fig3}(c). With $Q_c\approx3800$, the resonator is undercoupled in the entire range. The periodic dependence in $Q_i$ shows that the loss is maximal at $\varphi=(2n+1)\pi$, minimal at $\varphi = 2n\pi$.
We interpret the variations in $Q_i$ qualitatively within the ABS framework as follows. Each of the open transmission modes of the normal region results in a pair of bound states with energy dispersions $\pm E_n(\varphi)$ symmetric around $E=0$. With many partially open modes, a dense spectrum is formed with a minigap which is smallest at $\varphi=\pi$ as illustrated in Fig.\,\ref{fig3}(d). Consistently with theoretical expectations\,\cite{Virtanen2011} and experimental observations in similar systems\,\cite{ Dassonneville2018, Haller2022}, the dissipation in the JJ is most prominent around this phase, which reduces the resonator Q factor.

We note that the Al$_2$O$_3$ layer covers the entire chip, including the resonator. In bare resonators fabricated similarly on the same type of wafer, but without the SQUID and Al$_2$O$_3$ layer, we measured quality factors up to $Q_i=1.5\cdot10^4$. Therefore, we attribute the relatively high loss to the dissipation in the two-level fluctuators of the Al$_2$O$_3$ dielectric layer\,\cite{Corey2020}.

In the measurements discussed so far and plotted in Fig.\,\ref{fig3}(a-c), the 2DEG in the Josephson junction had a relatively high electron density at an applied gate voltage of $V_g=-7\,$V, up to $n \sim 1\cdot 10^{12}~\mathrm{cm^{-2}}$ based on Ref.\,\citenum{Suto2022}. To investigate how the CPR evolves with $V_g$, we repeated the procedure described above in a wide gate range. The gate dependence of the CPR Fourier coefficients $A_k$ are presented in Fig.\,\ref{fig3}(e). The JJ is completely depleted at $V_g=-10\,$V, and the onset of the transport is around $V_g=-9\,$V. We note that the gate-dependence was hysteretic and the threshold varied between measurements. The supercurrent is maximal near $V_G=-7\,$V and, interestingly, slightly decreases for more positive gate voltages, then remains approximately constant. Similar behavior has been observed in Ref.\,\cite{Suto2022} and is attributed to the opening of a second subband of the 2DEG. In the same panel we plot the effective channel transparency parameter $\tau$, which we determine by fitting the CPR curves with the short junction formula in the zero-temperature limit,
$I_s (\varphi) \propto {\tau \sin(\varphi)}/{\sqrt{1-\tau \sin ^2(\varphi/2)}}$.
The magnitude of the supercurrent indicates that there are multiple channels, nevertheless $\tau$ is a good metric to characterize the skewness of the CPR. 
The extracted $\tau$ also assumes its maximal value near $V_G=-7\,$V, as shown in Fig.\,\ref{fig3}(e). 

In the following, we discuss the characteristic length scales of the transport in the InAs section of the JJ. In a similar device investigated by DC measurements\,\cite{Suto2022} at the maximal critical current a charge carrier density of $n=6.5\times10^{11}\,\textrm{cm}^{-2}$, mean free path $l=200\,$nm and superconducting gap $\Delta=125\,$\textmu eV were estimated. The clean (ballistic) coherence length is $\xi = \hbar v_F/\pi \Delta=1.4\,$\textmu m with $v_F = \hbar \sqrt{2\pi n}/0.028 m_e$, and the diffusive coherence length is $\xi_d = \sqrt{\xi l}=530\,$nm. That is, at the maximal critical current $\xi_d \sim L$, the JJ is in the intermediate regime. Towards pinchoff, as $n$ decreases, so does $\xi_d$ and the JJ goes into the long limit.

\subsection{Supercurrent interference in out-of-plane magnetic field}

\begin{figure}[htb]
\includegraphics{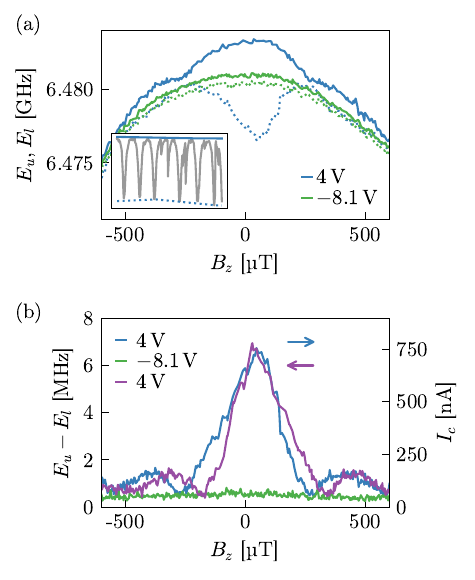}
\caption{\label{FigFraunhofer} Supercurrent interference measurement (device A). (a) Upper envelope $E_u$ (solid) and lower envelope $E_l$ (dotted) of the resonant frequency $f_0$, as a function of the out-of-plane magnetic field $B_z$. The measurement was performed at two gate voltages, $V_g=-8.1\,$V (junction in depletion) and $V_g=4\,$V (junction is open). As an example for the envelope determination, the inset shows $f_0$ data points (gray) in a small range from $B_z=0$ to $25\,$\textmu T, together with the envelopes (same color coding as in the main plot). (b) Difference of the upper and lower envelopes $\delta E$. The right axis shows the critical current $I_c$ inferred from $\delta E$. At $V_g=4\,$V, blue and purple lines indicate opposite $B_z$ sweep directions (shown by arrows). }
\end{figure}

We have investigated the supercurrent interference effect by applying the out-of-plane magnetic field on a larger scale, up to $|B_z|=600\,$\textmu T. For a quantitative evaluation we determined the upper and lower envelopes $E_{u,l}$ of the $f_0(B_z)$ resonance frequency (see the SM for details) and plotted them in Fig.\,\ref{FigFraunhofer}(a) in solid and dotted lines, respectively.

With the junction in depletion at $V_g=-8.1\,$V (green curves), the envelopes follow a near-parabolic function, with the maximum around $B_z=0$.
This $E_{u,l} \propto -B_z^2$ dependence is the result of the changing kinetic inductance of the resonator due to the pair-breaking effect of the magnetic field\,\cite{Healey2008, Samkharadze2016}. The curve is slightly skewed, not perfectly symmetric. We have observed that with an opposite sweep direction of the magnetic field, the skewness appears mirrored, and the curve shows hysteresis. While bulk Al is a Type-I superconductor, thin films have been shown to be Type-II\,\cite{Lopez-Nunez2023} and thus it can host vortices. We attribute the skewness and its hysteretic behavior to vortices forming in the epitaxial Al layer which forms the resonator\,\cite{Bothner2012}. 

When the junction is opened and tuned to the saturation regime by applying $V_g=4\,$V, the envelopes exhibit a more complex response (blue curves in Fig.\,\ref{FigFraunhofer}(a)). On the $\sim$\textmu T scale, the resonance frequency $f_0$ shows the periodic variation originating from the CPR as shown by the inset in Fig.\,\ref{FigFraunhofer}(a). This is the same deterministic response as in Fig.\,\ref{fig3}(a). In addition to this function, we observe random jumps, also exemplified in the inset, which get more frequent at higher magnetic fields. We attribute these jumps to external flux noise and vortex dynamics in the Al layer.
On the $\sim100\,$\textmu T scale, two features can be observed: an overall parabolic dependence originating from the kinetic inductance change of the resonator, and an oscillating modulation of the envelopes.
To evaluate the latter behaviour quantitatively, we define the envelope size \begin{equation}
\label{eq:envelope_size}
\delta E (B_z) = E_u (B_z) - E_l (B_z).
\end{equation}
and plot it in Fig.\,\ref{FigFraunhofer}(b) for three cases. With the depleted junction the difference is constant to a good approximation (green). With the junction measured at $V_g=4\,$V (blue and purple), a distinct pattern can be observed, which originates from supercurrent interference in the junction. The main lobe of the pattern exhibits skewness, which changes sign as the sweep direction is reversed (indicated with arrows).
We estimate the critical current by assuming a sinusoidal current-phase relation, $I_s = I_c \sin \varphi$, and matching the envelope size with the maximal frequency shift
\begin{equation}
\label{eq:maxshift}
\delta E  =  f_0 (\varphi=0) - f_0(\varphi=\pi)
\end{equation}
in our circuit model as a function of $I_c$. The critical current $I_c$ calculated from the envelope size is shown on the right axis in Fig.\,\ref{FigFraunhofer}(b). We interpret the green curve, originating from the evaluation of the unloaded resonator, as the noise floor of this method, giving a detection limit of $I_c^\mathrm{min}\approx50\,$nA. Theoretically, in the case of destructive interference we expect a total suppression of the supercurrent, however, we cannot resolve the minima of the interference pattern lower than this value.

In the interference pattern the destructive interference manifests in nodes at magnetic fields $B_n = n\Delta B_z$ for $n\in \mathbb{Z}$, with $\Delta B_z=\Phi_0/A$, where $A=LW$ is the JJ area.
We calculate the effective JJ area and the corresponding node spacing $\Delta B_z$ with taking into account the magnetic penetration in the superconducting leads as follows. In clean, bulk Al the London penetration depth is $\lambda_{\rm Al}=16\,$nm, and the coherence length is $\xi_{\rm Al}=1.6\,$\textmu m\,\cite{Merservey1969}.
The London penetration depth in impure Al is $\lambda = \lambda_{\rm Al} \sqrt{1+\xi_{\rm Al}/l}$, where $l$ is the mean free path, which we approximate with the film thickness $t=50\,$nm as an upper limit and get $\lambda \approx 92\,$nm. The penetration depth in the thin film is the Pearl length $\Lambda=\lambda\coth(t/\lambda)\approx 185\,$nm\,\cite{Pearl1964}.
The magnetic field penetrates the Al leads on this scale and thus the effective length of the JJ is $L_{\rm eff} = L + 2 \Lambda$. Without taking the shielding into account, the expected periodicity is $\Delta B_z = \Phi_0/W(L+2\Lambda)\approx 406\,$\textmu T.  
Experimentally, we obtain $\Delta B_z\approx260\,$\textmu T by taking half of the node-to-node distance of the main lobe. Taking into account the magnetic shielding as well, the $\sim 8$-fold discrepancy may be explained by an overestimation of the mean free path $l$ and flux focusing\,\cite{Zeldov1994}: the magnetic flux expelled from the Al layer is focused in the JJ area, making the flux higher than the uniform external flux and reducing the node spacing.

\subsection{In-plane magnetic field}

The application of an in-plane magnetic field and the presence of SOC can lead to novel effects from $\varphi_0$-junctions\,\cite{Szombati2016, Strambini2020, Mayer2020}, superconducting diode effects\,\cite{Lotfizadeh2024, Baumgartner2021, Reinhardt2024} to topological phases\,\cite{Mourik2012, Sarma2015, Deng2016, Aguado2017, Fornieri2019, Dartiailh2021}. Therefore, we repeated the interference experiment under the application of an in-plane magnetic field with varying direction and magnitude. We performed this in device B, nominally identical to device A, without magnetic shielding around the sample. 

\begin{figure}[htb]
\includegraphics{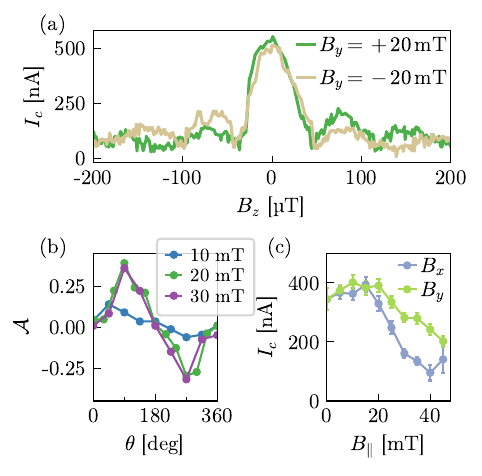}
\caption{\label{Fig5} Effects of the in-plane magnetic field (device B). (a) Supercurrent interference pattern in $B_y=\pm20\,$mT. The first side peaks have different height, the asymmetry is mirrored upon reversing the in-plane field. (b) $B_\|$ magnitude and angle dependence of the asymmetry ($\theta$ is the angle between $B_\|$ and the $x$-axis). (c) Suppression of the critical current with in-plane magnetic field ($B_z=0$) along two directions, $B_\|=B_x$ ($\theta=0$\textdegree) and $B_\|=B_y$ ($\theta=90$\textdegree).}
\end{figure}

In Fig.\,\ref{Fig5}(a) we plot two interference patterns measured with the in-plane magnetic field perpendicular to the current flow, in $B_y=\pm20\,$mT (corresponding to $\theta=90$\textdegree and $\theta=270$\textdegree, see also Fig.\,\ref{fig:sample}(c)). The measurement was evaluated as presented in the previous section, by determining the envelopes of the resonance shift, and converting the envelope size to current. Surprisingly, the first side lobes appear with different current magnitudes, and the asymmetry is mirrored by reversing the direction of the in-plane field. Compared to the interference pattern in Fig.\,\ref{FigFraunhofer}, there is a $\sim5$-fold difference in the magnetic field scale of the pattern, which we attribute to the lack of magnetic shielding. 

We quantify the asymmetry through the amplitude of the first side lobes in the interference pattern with
\begin{equation}
    \mathcal{A}=\frac{I_c^{(1)}-I_c^{(-1)}}{I_c^{(1)}+I_c^{(-1)}},
\end{equation}
where $I_c^{(\pm1)}$ are the maxima of the first right/left lobes\,\cite{Suominen2017}. We show the asymmetry dependence on the direction and magnitude of the in-plane field in Fig.\,\ref{Fig5}(b). In $B_\|=10$\,mT, there is a modest asymmetry $|\mathcal{A}|<0.15$, which grows as high as $|\mathcal{A}|\approx0.33$ in $B_\|=20$\,mT at $\theta = 90$\textdegree. $|\mathcal{A}|$ appears maximal (minimal) when the in-plane field is perpendicular (parallel) to the direction of the current.
Interestingly, the asymmetry does not increase when going from $B_\|=20$\,mT to $30$\,mT. 
It has been observed that in the presence of SOC and in-plane magnetic field the side peaks in the interference pattern exhibit asymmetry\,\cite{Suominen2017, Assouline2019}.
If the in-plane magnetic field is perpendicular to the current (and the out-of-plane Rashba field direction), the CPR exhibits a $\varphi_0$ phase shift.
This alone does not create an asymmetry in the interference pattern, however, if the $\varphi_0$ phase is spatially dependent (perpendicular to the current direction), e.g. due to disorder in the SOC strength, then it leads to the observed asymmetry in the interference pattern.
The angle dependence, i.e. maximum at 90\textdegree~and 270\textdegree~in $|\mathcal{A}|$ is consistent with the appearance of the largest $\varphi_0$ shift at these angles.

In Fig.\,\ref{Fig5}(c) we plot the critical current as the function of the in-plane magnetic field for two directions, along the $x$ ($\theta=0$\textdegree) and $y$ axis ($\theta=90$\textdegree).
In both cases, $I_c$ is suppressed by a factor of $\sim2$ in 45\,mT. The suppression of critical current in the $y$-direction have been investigated in several works.
The suppression is often followed by a revival of the critical current at higher field and is also accompanied by closing and reopening of the gap\,\cite{Fornieri2019, Haxell2023, Banerjee2023, Dartiailh2021}.
This is attributed to a $0-\pi$ transition in the junction. The topological origin of this is unclear, since orbital effects in the region below the contacts might also play a role\,\cite{Haxell2023}. In our case, we could not resolve the critical current for higher fields. In future devices better shielding and resonators that work up to higher magnetic fields are needed.
Finally, the stronger suppression in the $x$-direction is attributed to the inhomogeneous magnetic field profile due to the Meissner effect of the contacts leading to the formation of flux dipoles\,\cite{Suominen2017}. 

\section{\label{sec:conc}Conclusions}

Using a coplanar waveguide resonator, we have experimentally determined the current-phase relation in a Josephson junction where the weak link is the 2DEG formed in an InAs/InGaAs heterostructure. 
Furthermore, we have observed supercurrent interference in an out-of-plane magnetic field via the microwave signal, and studied the asymmetry of the interference pattern which arises under the application of an in-plane field. 
These results were obtained in spite of the relatively low quality factor of the resonator, which we attribute to the Al$_2$O$_3$ dielectric layer covering the whole chip. For narrower junctions with a smaller critical current, a larger sensitivity is required, therefore in future devices the dielectric will be deposited only on the gated region.
This is expected to yield a higher quality factor, enabling better sensitivity to study single-mode junctions and the direct mapping of Andreev states using two-tone spectroscopy.

\section{\label{sec:acknowledgments}Acknowledgments}
This work acknowledges support from the MultiSpin and 2DSOTECH FlagERA networks, the OTKA K138433, K134437 and PD134758 grants, the VEKOP 2.3.3-15-2017-00015 grant, the EIC Pathfinder Challenge grant QuKiT, the EU SuperGate network, the COST Action CA21144 (SUPERQUMAT), CA20116 (OPERA), PNRR MUR project PE0000023-NQSTI, and the EU Horizon 2020 programme under grant agreement No 101007417 within the framework of the NFFA-Europe Pilot Transnational Access Activity (proposal ID 380 in collaboration with CNR, Trieste). This research was supported by the Ministry of Culture and Innovation and the National Research, Development and Innovation Office within the Quantum Information National Laboratory of Hungary (Grant No. 2022-2.1.1-NL-2022-00004) and the Doctoral Excellence Fellowship Programme (DCEP), by the FET Open AndQC and SuperGate networks and by the European Research Council ERC project Twistrain, the János Bolyai Research Scholarship of the Hungarian Academy of Sciences, and the UNKP-23-5 New National Excellence Program.

The heterostructures were grown by M.K. and G.B., the fabrication was done by M. S., B.V., M. B. and E.T. The measurements and the data analysis was performed by Z. S., D. K., Cs. H., T. K. and G. F.. The project was guided by G.F., E.T., P.M. and Sz. Cs.. The manuscript has been prepared by G.F., M. S. and P.M. with input from all authors.
\bibliography{library}

\begin{thebibliography}{73}%
\makeatletter
\providecommand \@ifxundefined [1]{%
 \@ifx{#1\undefined}
}%
\providecommand \@ifnum [1]{%
 \ifnum #1\expandafter \@firstoftwo
 \else \expandafter \@secondoftwo
 \fi
}%
\providecommand \@ifx [1]{%
 \ifx #1\expandafter \@firstoftwo
 \else \expandafter \@secondoftwo
 \fi
}%
\providecommand \natexlab [1]{#1}%
\providecommand \enquote  [1]{``#1''}%
\providecommand \bibnamefont  [1]{#1}%
\providecommand \bibfnamefont [1]{#1}%
\providecommand \citenamefont [1]{#1}%
\providecommand \href@noop [0]{\@secondoftwo}%
\providecommand \href [0]{\begingroup \@sanitize@url \@href}%
\providecommand \@href[1]{\@@startlink{#1}\@@href}%
\providecommand \@@href[1]{\endgroup#1\@@endlink}%
\providecommand \@sanitize@url [0]{\catcode `\\12\catcode `\$12\catcode `\&12\catcode `\#12\catcode `\^12\catcode `\_12\catcode `\%12\relax}%
\providecommand \@@startlink[1]{}%
\providecommand \@@endlink[0]{}%
\providecommand \url  [0]{\begingroup\@sanitize@url \@url }%
\providecommand \@url [1]{\endgroup\@href {#1}{\urlprefix }}%
\providecommand \urlprefix  [0]{URL }%
\providecommand \Eprint [0]{\href }%
\providecommand \doibase [0]{https://doi.org/}%
\providecommand \selectlanguage [0]{\@gobble}%
\providecommand \bibinfo  [0]{\@secondoftwo}%
\providecommand \bibfield  [0]{\@secondoftwo}%
\providecommand \translation [1]{[#1]}%
\providecommand \BibitemOpen [0]{}%
\providecommand \bibitemStop [0]{}%
\providecommand \bibitemNoStop [0]{.\EOS\space}%
\providecommand \EOS [0]{\spacefactor3000\relax}%
\providecommand \BibitemShut  [1]{\csname bibitem#1\endcsname}%
\let\auto@bib@innerbib\@empty
\bibitem [{\citenamefont {Kjaergaard}\ \emph {et~al.}(2020)\citenamefont {Kjaergaard}, \citenamefont {Schwartz}, \citenamefont {Braumüller}, \citenamefont {Krantz}, \citenamefont {Wang}, \citenamefont {Gustavsson},\ and\ \citenamefont {Oliver}}]{Kjaergaard2020}%
  \BibitemOpen
  \bibfield  {author} {\bibinfo {author} {\bibfnamefont {M.}~\bibnamefont {Kjaergaard}}, \bibinfo {author} {\bibfnamefont {M.~E.}\ \bibnamefont {Schwartz}}, \bibinfo {author} {\bibfnamefont {J.}~\bibnamefont {Braumüller}}, \bibinfo {author} {\bibfnamefont {P.}~\bibnamefont {Krantz}}, \bibinfo {author} {\bibfnamefont {J.~I.}\ \bibnamefont {Wang}}, \bibinfo {author} {\bibfnamefont {S.}~\bibnamefont {Gustavsson}},\ and\ \bibinfo {author} {\bibfnamefont {W.~D.}\ \bibnamefont {Oliver}},\ }\href {https://doi.org/10.1146/annurev-conmatphys-031119-050605} {\bibfield  {journal} {\bibinfo  {journal} {Annual Review of Condensed Matter Physics}\ }\textbf {\bibinfo {volume} {11}},\ \bibinfo {pages} {369} (\bibinfo {year} {2020})}\BibitemShut {NoStop}%
\bibitem [{\citenamefont {Larsen}\ \emph {et~al.}(2015)\citenamefont {Larsen}, \citenamefont {Petersson}, \citenamefont {Kuemmeth}, \citenamefont {Jespersen}, \citenamefont {Krogstrup}, \citenamefont {Nyg\aa{}rd},\ and\ \citenamefont {Marcus}}]{PhysRevLett.115.127001}%
  \BibitemOpen
  \bibfield  {author} {\bibinfo {author} {\bibfnamefont {T.~W.}\ \bibnamefont {Larsen}}, \bibinfo {author} {\bibfnamefont {K.~D.}\ \bibnamefont {Petersson}}, \bibinfo {author} {\bibfnamefont {F.}~\bibnamefont {Kuemmeth}}, \bibinfo {author} {\bibfnamefont {T.~S.}\ \bibnamefont {Jespersen}}, \bibinfo {author} {\bibfnamefont {P.}~\bibnamefont {Krogstrup}}, \bibinfo {author} {\bibfnamefont {J.}~\bibnamefont {Nyg\aa{}rd}},\ and\ \bibinfo {author} {\bibfnamefont {C.~M.}\ \bibnamefont {Marcus}},\ }\href {https://doi.org/10.1103/PhysRevLett.115.127001} {\bibfield  {journal} {\bibinfo  {journal} {Phys. Rev. Lett.}\ }\textbf {\bibinfo {volume} {115}},\ \bibinfo {pages} {127001} (\bibinfo {year} {2015})}\BibitemShut {NoStop}%
\bibitem [{\citenamefont {Strickland}\ \emph {et~al.}(2024{\natexlab{a}})\citenamefont {Strickland}, \citenamefont {Baker}, \citenamefont {Lee}, \citenamefont {Dindial}, \citenamefont {Elfeky}, \citenamefont {Strohbeen}, \citenamefont {Hatefipour}, \citenamefont {Yu}, \citenamefont {Levy}, \citenamefont {Issokson}, \citenamefont {Manucharyan},\ and\ \citenamefont {Shabani}}]{PhysRevResearch.6.023094}%
  \BibitemOpen
  \bibfield  {author} {\bibinfo {author} {\bibfnamefont {W.~M.}\ \bibnamefont {Strickland}}, \bibinfo {author} {\bibfnamefont {L.~J.}\ \bibnamefont {Baker}}, \bibinfo {author} {\bibfnamefont {J.}~\bibnamefont {Lee}}, \bibinfo {author} {\bibfnamefont {K.}~\bibnamefont {Dindial}}, \bibinfo {author} {\bibfnamefont {B.~H.}\ \bibnamefont {Elfeky}}, \bibinfo {author} {\bibfnamefont {P.~J.}\ \bibnamefont {Strohbeen}}, \bibinfo {author} {\bibfnamefont {M.}~\bibnamefont {Hatefipour}}, \bibinfo {author} {\bibfnamefont {P.}~\bibnamefont {Yu}}, \bibinfo {author} {\bibfnamefont {I.}~\bibnamefont {Levy}}, \bibinfo {author} {\bibfnamefont {J.}~\bibnamefont {Issokson}}, \bibinfo {author} {\bibfnamefont {V.~E.}\ \bibnamefont {Manucharyan}},\ and\ \bibinfo {author} {\bibfnamefont {J.}~\bibnamefont {Shabani}},\ }\href {https://doi.org/10.1103/PhysRevResearch.6.023094} {\bibfield  {journal} {\bibinfo  {journal} {Phys. Rev. Res.}\ }\textbf {\bibinfo {volume} {6}},\ \bibinfo {pages} {023094} (\bibinfo {year}
  {2024}{\natexlab{a}})}\BibitemShut {NoStop}%
\bibitem [{\citenamefont {de~Lange}\ \emph {et~al.}(2015)\citenamefont {de~Lange}, \citenamefont {van Heck}, \citenamefont {Bruno}, \citenamefont {van Woerkom}, \citenamefont {Geresdi}, \citenamefont {Plissard}, \citenamefont {Bakkers}, \citenamefont {Akhmerov},\ and\ \citenamefont {DiCarlo}}]{PhysRevLett.115.127002}%
  \BibitemOpen
  \bibfield  {author} {\bibinfo {author} {\bibfnamefont {G.}~\bibnamefont {de~Lange}}, \bibinfo {author} {\bibfnamefont {B.}~\bibnamefont {van Heck}}, \bibinfo {author} {\bibfnamefont {A.}~\bibnamefont {Bruno}}, \bibinfo {author} {\bibfnamefont {D.~J.}\ \bibnamefont {van Woerkom}}, \bibinfo {author} {\bibfnamefont {A.}~\bibnamefont {Geresdi}}, \bibinfo {author} {\bibfnamefont {S.~R.}\ \bibnamefont {Plissard}}, \bibinfo {author} {\bibfnamefont {E.~P. A.~M.}\ \bibnamefont {Bakkers}}, \bibinfo {author} {\bibfnamefont {A.~R.}\ \bibnamefont {Akhmerov}},\ and\ \bibinfo {author} {\bibfnamefont {L.}~\bibnamefont {DiCarlo}},\ }\href {https://doi.org/10.1103/PhysRevLett.115.127002} {\bibfield  {journal} {\bibinfo  {journal} {Phys. Rev. Lett.}\ }\textbf {\bibinfo {volume} {115}},\ \bibinfo {pages} {127002} (\bibinfo {year} {2015})}\BibitemShut {NoStop}%
\bibitem [{\citenamefont {Casparis}\ \emph {et~al.}(2018)\citenamefont {Casparis}, \citenamefont {Connolly}, \citenamefont {Kjaergaard}, \citenamefont {Pearson}, \citenamefont {Kringhøj}, \citenamefont {Larsen}, \citenamefont {Kuemmeth}, \citenamefont {Wang}, \citenamefont {Thomas}, \citenamefont {Gronin}, \citenamefont {Gardner}, \citenamefont {Manfra}, \citenamefont {Marcus},\ and\ \citenamefont {Petersson}}]{Casparis2018}%
  \BibitemOpen
  \bibfield  {author} {\bibinfo {author} {\bibfnamefont {L.}~\bibnamefont {Casparis}}, \bibinfo {author} {\bibfnamefont {M.~R.}\ \bibnamefont {Connolly}}, \bibinfo {author} {\bibfnamefont {M.}~\bibnamefont {Kjaergaard}}, \bibinfo {author} {\bibfnamefont {N.~J.}\ \bibnamefont {Pearson}}, \bibinfo {author} {\bibfnamefont {A.}~\bibnamefont {Kringhøj}}, \bibinfo {author} {\bibfnamefont {T.~W.}\ \bibnamefont {Larsen}}, \bibinfo {author} {\bibfnamefont {F.}~\bibnamefont {Kuemmeth}}, \bibinfo {author} {\bibfnamefont {T.}~\bibnamefont {Wang}}, \bibinfo {author} {\bibfnamefont {C.}~\bibnamefont {Thomas}}, \bibinfo {author} {\bibfnamefont {S.}~\bibnamefont {Gronin}}, \bibinfo {author} {\bibfnamefont {G.~C.}\ \bibnamefont {Gardner}}, \bibinfo {author} {\bibfnamefont {M.~J.}\ \bibnamefont {Manfra}}, \bibinfo {author} {\bibfnamefont {C.~M.}\ \bibnamefont {Marcus}},\ and\ \bibinfo {author} {\bibfnamefont {K.~D.}\ \bibnamefont {Petersson}},\ }\href {https://doi.org/10.1038/s41565-018-0207-y} {\bibfield  {journal} {\bibinfo
   {journal} {Nature Nanotechnology}\ }\textbf {\bibinfo {volume} {13}},\ \bibinfo {pages} {915} (\bibinfo {year} {2018})}\BibitemShut {NoStop}%
\bibitem [{\citenamefont {Zheng}\ \emph {et~al.}(2023)\citenamefont {Zheng}, \citenamefont {Cheung}, \citenamefont {Sangwan}, \citenamefont {Kononov}, \citenamefont {Haller}, \citenamefont {Ridderbos}, \citenamefont {Ciaccia}, \citenamefont {Ungerer}, \citenamefont {Li}, \citenamefont {Bakkers}, \citenamefont {Baumgartner},\ and\ \citenamefont {Sch{\"o}nenberger}}]{Zheng2023CoherentCO}%
  \BibitemOpen
  \bibfield  {author} {\bibinfo {author} {\bibfnamefont {H.}~\bibnamefont {Zheng}}, \bibinfo {author} {\bibfnamefont {L.~Y.}\ \bibnamefont {Cheung}}, \bibinfo {author} {\bibfnamefont {N.}~\bibnamefont {Sangwan}}, \bibinfo {author} {\bibfnamefont {A.}~\bibnamefont {Kononov}}, \bibinfo {author} {\bibfnamefont {R.}~\bibnamefont {Haller}}, \bibinfo {author} {\bibfnamefont {J.}~\bibnamefont {Ridderbos}}, \bibinfo {author} {\bibfnamefont {C.}~\bibnamefont {Ciaccia}}, \bibinfo {author} {\bibfnamefont {J.~H.}\ \bibnamefont {Ungerer}}, \bibinfo {author} {\bibfnamefont {A.}~\bibnamefont {Li}}, \bibinfo {author} {\bibfnamefont {E.~P.}\ \bibnamefont {Bakkers}}, \bibinfo {author} {\bibfnamefont {A.}~\bibnamefont {Baumgartner}},\ and\ \bibinfo {author} {\bibfnamefont {C.}~\bibnamefont {Sch{\"o}nenberger}},\ }\href {https://pubs.acs.org/doi/10.1021/acs.nanolett.4c00770} {\bibfield  {journal} {\bibinfo  {journal} {Nano Letters}\ }\textbf {\bibinfo {volume} {24}},\ \bibinfo {pages} {7173 } (\bibinfo {year}
  {2023})}\BibitemShut {NoStop}%
\bibitem [{\citenamefont {Sagi}\ \emph {et~al.}(2024)\citenamefont {Sagi}, \citenamefont {Crippa}, \citenamefont {Valentini}, \citenamefont {Janik}, \citenamefont {Baghumyan}, \citenamefont {Fabris}, \citenamefont {Kapoor}, \citenamefont {Hassani}, \citenamefont {Fink}, \citenamefont {Calcaterra}, \citenamefont {Chrastina}, \citenamefont {Isella},\ and\ \citenamefont {Katsaros}}]{Sagi2024}%
  \BibitemOpen
  \bibfield  {author} {\bibinfo {author} {\bibfnamefont {O.}~\bibnamefont {Sagi}}, \bibinfo {author} {\bibfnamefont {A.}~\bibnamefont {Crippa}}, \bibinfo {author} {\bibfnamefont {M.}~\bibnamefont {Valentini}}, \bibinfo {author} {\bibfnamefont {M.}~\bibnamefont {Janik}}, \bibinfo {author} {\bibfnamefont {L.}~\bibnamefont {Baghumyan}}, \bibinfo {author} {\bibfnamefont {G.}~\bibnamefont {Fabris}}, \bibinfo {author} {\bibfnamefont {L.}~\bibnamefont {Kapoor}}, \bibinfo {author} {\bibfnamefont {F.}~\bibnamefont {Hassani}}, \bibinfo {author} {\bibfnamefont {J.}~\bibnamefont {Fink}}, \bibinfo {author} {\bibfnamefont {S.}~\bibnamefont {Calcaterra}}, \bibinfo {author} {\bibfnamefont {D.}~\bibnamefont {Chrastina}}, \bibinfo {author} {\bibfnamefont {G.}~\bibnamefont {Isella}},\ and\ \bibinfo {author} {\bibfnamefont {G.}~\bibnamefont {Katsaros}},\ }\href {https://arxiv.org/abs/2403.16774v1} {\bibinfo {title} {A gate tunable transmon qubit in planar {G}e}} (\bibinfo {year} {2024})\BibitemShut {NoStop}%
\bibitem [{\citenamefont {Hays}\ \emph {et~al.}(2020)\citenamefont {Hays}, \citenamefont {Fatemi}, \citenamefont {Serniak}, \citenamefont {Bouman}, \citenamefont {Diamond}, \citenamefont {de~Lange}, \citenamefont {Krogstrup}, \citenamefont {Nygård}, \citenamefont {Geresdi},\ and\ \citenamefont {Devoret}}]{Hays2020}%
  \BibitemOpen
  \bibfield  {author} {\bibinfo {author} {\bibfnamefont {M.}~\bibnamefont {Hays}}, \bibinfo {author} {\bibfnamefont {V.}~\bibnamefont {Fatemi}}, \bibinfo {author} {\bibfnamefont {K.}~\bibnamefont {Serniak}}, \bibinfo {author} {\bibfnamefont {D.}~\bibnamefont {Bouman}}, \bibinfo {author} {\bibfnamefont {S.}~\bibnamefont {Diamond}}, \bibinfo {author} {\bibfnamefont {G.}~\bibnamefont {de~Lange}}, \bibinfo {author} {\bibfnamefont {P.}~\bibnamefont {Krogstrup}}, \bibinfo {author} {\bibfnamefont {J.}~\bibnamefont {Nygård}}, \bibinfo {author} {\bibfnamefont {A.}~\bibnamefont {Geresdi}},\ and\ \bibinfo {author} {\bibfnamefont {M.~H.}\ \bibnamefont {Devoret}},\ }\href {https://doi.org/10.1038/s41567-020-0952-3} {\bibfield  {journal} {\bibinfo  {journal} {Nature Physics 2020 16:11}\ }\textbf {\bibinfo {volume} {16}},\ \bibinfo {pages} {1103} (\bibinfo {year} {2020})}\BibitemShut {NoStop}%
\bibitem [{\citenamefont {Hays}\ \emph {et~al.}(2021)\citenamefont {Hays}, \citenamefont {Fatemi}, \citenamefont {Bouman}, \citenamefont {Cerrillo}, \citenamefont {Diamond}, \citenamefont {Serniak}, \citenamefont {Connolly}, \citenamefont {Krogstrup}, \citenamefont {Nygård}, \citenamefont {Yeyati}, \citenamefont {Geresdi},\ and\ \citenamefont {Devoret}}]{Hays2021}%
  \BibitemOpen
  \bibfield  {author} {\bibinfo {author} {\bibfnamefont {M.}~\bibnamefont {Hays}}, \bibinfo {author} {\bibfnamefont {V.}~\bibnamefont {Fatemi}}, \bibinfo {author} {\bibfnamefont {D.}~\bibnamefont {Bouman}}, \bibinfo {author} {\bibfnamefont {J.}~\bibnamefont {Cerrillo}}, \bibinfo {author} {\bibfnamefont {S.}~\bibnamefont {Diamond}}, \bibinfo {author} {\bibfnamefont {K.}~\bibnamefont {Serniak}}, \bibinfo {author} {\bibfnamefont {T.}~\bibnamefont {Connolly}}, \bibinfo {author} {\bibfnamefont {P.}~\bibnamefont {Krogstrup}}, \bibinfo {author} {\bibfnamefont {J.}~\bibnamefont {Nygård}}, \bibinfo {author} {\bibfnamefont {A.~L.}\ \bibnamefont {Yeyati}}, \bibinfo {author} {\bibfnamefont {A.}~\bibnamefont {Geresdi}},\ and\ \bibinfo {author} {\bibfnamefont {M.~H.}\ \bibnamefont {Devoret}},\ }\href {https://doi.org/10.1126/science.abf0345} {\bibfield  {journal} {\bibinfo  {journal} {Science}\ }\textbf {\bibinfo {volume} {373}},\ \bibinfo {pages} {430} (\bibinfo {year} {2021})}\BibitemShut {NoStop}%
\bibitem [{\citenamefont {Pita-Vidal}\ \emph {et~al.}(2023)\citenamefont {Pita-Vidal}, \citenamefont {Bargerbos}, \citenamefont {Žitko}, \citenamefont {Splitthoff}, \citenamefont {Grünhaupt}, \citenamefont {Wesdorp}, \citenamefont {Liu}, \citenamefont {Kouwenhoven}, \citenamefont {Aguado}, \citenamefont {van Heck}, \citenamefont {Kou},\ and\ \citenamefont {Andersen}}]{Pita2023}%
  \BibitemOpen
  \bibfield  {author} {\bibinfo {author} {\bibfnamefont {M.}~\bibnamefont {Pita-Vidal}}, \bibinfo {author} {\bibfnamefont {A.}~\bibnamefont {Bargerbos}}, \bibinfo {author} {\bibfnamefont {R.}~\bibnamefont {Žitko}}, \bibinfo {author} {\bibfnamefont {L.~J.}\ \bibnamefont {Splitthoff}}, \bibinfo {author} {\bibfnamefont {L.}~\bibnamefont {Grünhaupt}}, \bibinfo {author} {\bibfnamefont {J.~J.}\ \bibnamefont {Wesdorp}}, \bibinfo {author} {\bibfnamefont {Y.}~\bibnamefont {Liu}}, \bibinfo {author} {\bibfnamefont {L.~P.}\ \bibnamefont {Kouwenhoven}}, \bibinfo {author} {\bibfnamefont {R.}~\bibnamefont {Aguado}}, \bibinfo {author} {\bibfnamefont {B.}~\bibnamefont {van Heck}}, \bibinfo {author} {\bibfnamefont {A.}~\bibnamefont {Kou}},\ and\ \bibinfo {author} {\bibfnamefont {C.~K.}\ \bibnamefont {Andersen}},\ }\href {https://doi.org/10.1038/s41567-023-02071-x} {\bibfield  {journal} {\bibinfo  {journal} {Nature Physics 2023 19:8}\ }\textbf {\bibinfo {volume} {19}},\ \bibinfo {pages} {1110} (\bibinfo {year}
  {2023})}\BibitemShut {NoStop}%
\bibitem [{\citenamefont {Pita-Vidal}\ \emph {et~al.}(2024)\citenamefont {Pita-Vidal}, \citenamefont {Wesdorp}, \citenamefont {Splitthoff}, \citenamefont {Bargerbos}, \citenamefont {Liu}, \citenamefont {Kouwenhoven},\ and\ \citenamefont {Andersen}}]{Pita2024}%
  \BibitemOpen
  \bibfield  {author} {\bibinfo {author} {\bibfnamefont {M.}~\bibnamefont {Pita-Vidal}}, \bibinfo {author} {\bibfnamefont {J.~J.}\ \bibnamefont {Wesdorp}}, \bibinfo {author} {\bibfnamefont {L.~J.}\ \bibnamefont {Splitthoff}}, \bibinfo {author} {\bibfnamefont {A.}~\bibnamefont {Bargerbos}}, \bibinfo {author} {\bibfnamefont {Y.}~\bibnamefont {Liu}}, \bibinfo {author} {\bibfnamefont {L.~P.}\ \bibnamefont {Kouwenhoven}},\ and\ \bibinfo {author} {\bibfnamefont {C.~K.}\ \bibnamefont {Andersen}},\ }\href {https://doi.org/10.1038/s41567-024-02497-x} {\bibfield  {journal} {\bibinfo  {journal} {Nature Physics 2024}\ ,\ \bibinfo {pages} {1}} (\bibinfo {year} {2024})}\BibitemShut {NoStop}%
\bibitem [{\citenamefont {Strickland}\ \emph {et~al.}(2024{\natexlab{b}})\citenamefont {Strickland}, \citenamefont {Elfeky}, \citenamefont {Baker}, \citenamefont {Maiani}, \citenamefont {Lee}, \citenamefont {Levy}, \citenamefont {Issokson}, \citenamefont {Vrajitoarea},\ and\ \citenamefont {Shabani}}]{Strickland2024}%
  \BibitemOpen
  \bibfield  {author} {\bibinfo {author} {\bibfnamefont {W.~M.}\ \bibnamefont {Strickland}}, \bibinfo {author} {\bibfnamefont {B.~H.}\ \bibnamefont {Elfeky}}, \bibinfo {author} {\bibfnamefont {L.}~\bibnamefont {Baker}}, \bibinfo {author} {\bibfnamefont {A.}~\bibnamefont {Maiani}}, \bibinfo {author} {\bibfnamefont {J.}~\bibnamefont {Lee}}, \bibinfo {author} {\bibfnamefont {I.}~\bibnamefont {Levy}}, \bibinfo {author} {\bibfnamefont {J.}~\bibnamefont {Issokson}}, \bibinfo {author} {\bibfnamefont {A.}~\bibnamefont {Vrajitoarea}},\ and\ \bibinfo {author} {\bibfnamefont {J.}~\bibnamefont {Shabani}},\ }\href {https://arxiv.org/abs/2406.09002v2} {\bibinfo {title} {Gatemonium: A voltage-tunable fluxonium}} (\bibinfo {year} {2024}{\natexlab{b}})\BibitemShut {NoStop}%
\bibitem [{\citenamefont {Mourik}\ \emph {et~al.}(2012)\citenamefont {Mourik}, \citenamefont {Zuo}, \citenamefont {Frolov}, \citenamefont {Plissard}, \citenamefont {Bakkers},\ and\ \citenamefont {Kouwenhoven}}]{Mourik2012}%
  \BibitemOpen
  \bibfield  {author} {\bibinfo {author} {\bibfnamefont {V.}~\bibnamefont {Mourik}}, \bibinfo {author} {\bibfnamefont {K.}~\bibnamefont {Zuo}}, \bibinfo {author} {\bibfnamefont {S.~M.}\ \bibnamefont {Frolov}}, \bibinfo {author} {\bibfnamefont {S.~R.}\ \bibnamefont {Plissard}}, \bibinfo {author} {\bibfnamefont {E.~P.}\ \bibnamefont {Bakkers}},\ and\ \bibinfo {author} {\bibfnamefont {L.~P.}\ \bibnamefont {Kouwenhoven}},\ }\href {https://doi.org/10.1126/SCIENCE.1222360} {\bibfield  {journal} {\bibinfo  {journal} {Science}\ }\textbf {\bibinfo {volume} {336}},\ \bibinfo {pages} {1003} (\bibinfo {year} {2012})}\BibitemShut {NoStop}%
\bibitem [{\citenamefont {Sarma}\ \emph {et~al.}(2015)\citenamefont {Sarma}, \citenamefont {Freedman},\ and\ \citenamefont {Nayak}}]{Sarma2015}%
  \BibitemOpen
  \bibfield  {author} {\bibinfo {author} {\bibfnamefont {S.~D.}\ \bibnamefont {Sarma}}, \bibinfo {author} {\bibfnamefont {M.}~\bibnamefont {Freedman}},\ and\ \bibinfo {author} {\bibfnamefont {C.}~\bibnamefont {Nayak}},\ }\href {https://doi.org/10.1038/npjqi.2015.1} {\bibfield  {journal} {\bibinfo  {journal} {npj Quantum Information 2015 1:1}\ }\textbf {\bibinfo {volume} {1}},\ \bibinfo {pages} {1} (\bibinfo {year} {2015})}\BibitemShut {NoStop}%
\bibitem [{\citenamefont {Deng}\ \emph {et~al.}(2016)\citenamefont {Deng}, \citenamefont {Vaitiekenas}, \citenamefont {Hansen}, \citenamefont {Danon}, \citenamefont {Leijnse}, \citenamefont {Flensberg}, \citenamefont {Nygård}, \citenamefont {Krogstrup},\ and\ \citenamefont {Marcus}}]{Deng2016}%
  \BibitemOpen
  \bibfield  {author} {\bibinfo {author} {\bibfnamefont {M.~T.}\ \bibnamefont {Deng}}, \bibinfo {author} {\bibfnamefont {S.}~\bibnamefont {Vaitiekenas}}, \bibinfo {author} {\bibfnamefont {E.~B.}\ \bibnamefont {Hansen}}, \bibinfo {author} {\bibfnamefont {J.}~\bibnamefont {Danon}}, \bibinfo {author} {\bibfnamefont {M.}~\bibnamefont {Leijnse}}, \bibinfo {author} {\bibfnamefont {K.}~\bibnamefont {Flensberg}}, \bibinfo {author} {\bibfnamefont {J.}~\bibnamefont {Nygård}}, \bibinfo {author} {\bibfnamefont {P.}~\bibnamefont {Krogstrup}},\ and\ \bibinfo {author} {\bibfnamefont {C.~M.}\ \bibnamefont {Marcus}},\ }\href {https://doi.org/10.1126/SCIENCE.AAF3961} {\bibfield  {journal} {\bibinfo  {journal} {Science}\ }\textbf {\bibinfo {volume} {354}},\ \bibinfo {pages} {1557} (\bibinfo {year} {2016})}\BibitemShut {NoStop}%
\bibitem [{\citenamefont {Aguado}(2017)}]{Aguado2017}%
  \BibitemOpen
  \bibfield  {author} {\bibinfo {author} {\bibfnamefont {R.}~\bibnamefont {Aguado}},\ }\href {https://doi.org/10.1393/NCR/I2017-10141-9} {\bibfield  {journal} {\bibinfo  {journal} {Rivista del Nuovo Cimento}\ }\textbf {\bibinfo {volume} {40}},\ \bibinfo {pages} {523} (\bibinfo {year} {2017})}\BibitemShut {NoStop}%
\bibitem [{\citenamefont {Fornieri}\ \emph {et~al.}(2019)\citenamefont {Fornieri}, \citenamefont {Whiticar}, \citenamefont {Setiawan}, \citenamefont {Portolés}, \citenamefont {Drachmann}, \citenamefont {Keselman}, \citenamefont {Gronin}, \citenamefont {Thomas}, \citenamefont {Wang}, \citenamefont {Kallaher}, \citenamefont {Gardner}, \citenamefont {Berg}, \citenamefont {Manfra}, \citenamefont {Stern}, \citenamefont {Marcus},\ and\ \citenamefont {Nichele}}]{Fornieri2019}%
  \BibitemOpen
  \bibfield  {author} {\bibinfo {author} {\bibfnamefont {A.}~\bibnamefont {Fornieri}}, \bibinfo {author} {\bibfnamefont {A.~M.}\ \bibnamefont {Whiticar}}, \bibinfo {author} {\bibfnamefont {F.}~\bibnamefont {Setiawan}}, \bibinfo {author} {\bibfnamefont {E.}~\bibnamefont {Portolés}}, \bibinfo {author} {\bibfnamefont {A.~C.~C.}\ \bibnamefont {Drachmann}}, \bibinfo {author} {\bibfnamefont {A.}~\bibnamefont {Keselman}}, \bibinfo {author} {\bibfnamefont {S.}~\bibnamefont {Gronin}}, \bibinfo {author} {\bibfnamefont {C.}~\bibnamefont {Thomas}}, \bibinfo {author} {\bibfnamefont {T.}~\bibnamefont {Wang}}, \bibinfo {author} {\bibfnamefont {R.}~\bibnamefont {Kallaher}}, \bibinfo {author} {\bibfnamefont {G.~C.}\ \bibnamefont {Gardner}}, \bibinfo {author} {\bibfnamefont {E.}~\bibnamefont {Berg}}, \bibinfo {author} {\bibfnamefont {M.~J.}\ \bibnamefont {Manfra}}, \bibinfo {author} {\bibfnamefont {A.}~\bibnamefont {Stern}}, \bibinfo {author} {\bibfnamefont {C.~M.}\ \bibnamefont {Marcus}},\ and\ \bibinfo {author}
  {\bibfnamefont {F.}~\bibnamefont {Nichele}},\ }\href {https://doi.org/10.1038/s41586-019-1068-8} {\bibfield  {journal} {\bibinfo  {journal} {Nature}\ }\textbf {\bibinfo {volume} {569}},\ \bibinfo {pages} {89} (\bibinfo {year} {2019})}\BibitemShut {NoStop}%
\bibitem [{\citenamefont {Dartiailh}\ \emph {et~al.}(2021)\citenamefont {Dartiailh}, \citenamefont {Mayer}, \citenamefont {Yuan}, \citenamefont {Wickramasinghe}, \citenamefont {Matos-Abiague}, \citenamefont {\ifmmode \check{Z}\else \v{Z}\fi{}uti\ifmmode~\acute{c}\else \'{c}\fi{}},\ and\ \citenamefont {Shabani}}]{Dartiailh2021}%
  \BibitemOpen
  \bibfield  {author} {\bibinfo {author} {\bibfnamefont {M.~C.}\ \bibnamefont {Dartiailh}}, \bibinfo {author} {\bibfnamefont {W.}~\bibnamefont {Mayer}}, \bibinfo {author} {\bibfnamefont {J.}~\bibnamefont {Yuan}}, \bibinfo {author} {\bibfnamefont {K.~S.}\ \bibnamefont {Wickramasinghe}}, \bibinfo {author} {\bibfnamefont {A.}~\bibnamefont {Matos-Abiague}}, \bibinfo {author} {\bibfnamefont {I.}~\bibnamefont {\ifmmode \check{Z}\else \v{Z}\fi{}uti\ifmmode~\acute{c}\else \'{c}\fi{}}},\ and\ \bibinfo {author} {\bibfnamefont {J.}~\bibnamefont {Shabani}},\ }\href {https://doi.org/10.1103/PhysRevLett.126.036802} {\bibfield  {journal} {\bibinfo  {journal} {Phys. Rev. Lett.}\ }\textbf {\bibinfo {volume} {126}},\ \bibinfo {pages} {036802} (\bibinfo {year} {2021})}\BibitemShut {NoStop}%
\bibitem [{\citenamefont {Amundsen}\ \emph {et~al.}(2024)\citenamefont {Amundsen}, \citenamefont {Linder}, \citenamefont {Robinson}, \citenamefont {Žutić},\ and\ \citenamefont {Banerjee}}]{Amundsen2024}%
  \BibitemOpen
  \bibfield  {author} {\bibinfo {author} {\bibfnamefont {M.}~\bibnamefont {Amundsen}}, \bibinfo {author} {\bibfnamefont {J.}~\bibnamefont {Linder}}, \bibinfo {author} {\bibfnamefont {J.~W.}\ \bibnamefont {Robinson}}, \bibinfo {author} {\bibfnamefont {I.}~\bibnamefont {Žutić}},\ and\ \bibinfo {author} {\bibfnamefont {N.}~\bibnamefont {Banerjee}},\ }\href {https://doi.org/10.1103/REVMODPHYS.96.021003/FIGURES/9/MEDIUM} {\bibfield  {journal} {\bibinfo  {journal} {Reviews of Modern Physics}\ }\textbf {\bibinfo {volume} {96}},\ \bibinfo {pages} {021003} (\bibinfo {year} {2024})}\BibitemShut {NoStop}%
\bibitem [{\citenamefont {Sau}\ and\ \citenamefont {Sarma}(2012)}]{Sau2012}%
  \BibitemOpen
  \bibfield  {author} {\bibinfo {author} {\bibfnamefont {J.~D.}\ \bibnamefont {Sau}}\ and\ \bibinfo {author} {\bibfnamefont {S.~D.}\ \bibnamefont {Sarma}},\ }\href {https://doi.org/10.1038/ncomms1966} {\bibfield  {journal} {\bibinfo  {journal} {Nature Communications 2012 3:1}\ }\textbf {\bibinfo {volume} {3}},\ \bibinfo {pages} {1} (\bibinfo {year} {2012})}\BibitemShut {NoStop}%
\bibitem [{\citenamefont {Leijnse}\ and\ \citenamefont {Flensberg}(2012)}]{Leijnse2012}%
  \BibitemOpen
  \bibfield  {author} {\bibinfo {author} {\bibfnamefont {M.}~\bibnamefont {Leijnse}}\ and\ \bibinfo {author} {\bibfnamefont {K.}~\bibnamefont {Flensberg}},\ }\href {https://doi.org/10.1103/PHYSREVB.86.134528} {\bibfield  {journal} {\bibinfo  {journal} {Physical Review B - Condensed Matter and Materials Physics}\ }\textbf {\bibinfo {volume} {86}},\ \bibinfo {pages} {134528} (\bibinfo {year} {2012})}\BibitemShut {NoStop}%
\bibitem [{\citenamefont {Zatelli}\ \emph {et~al.}(2023)\citenamefont {Zatelli}, \citenamefont {van Driel}, \citenamefont {Xu}, \citenamefont {Wang}, \citenamefont {Liu}, \citenamefont {Bordin}, \citenamefont {Roovers}, \citenamefont {Mazur}, \citenamefont {van Loo}, \citenamefont {Wolff}, \citenamefont {Bozkurt}, \citenamefont {Badawy}, \citenamefont {Gazibegovic}, \citenamefont {Bakkers}, \citenamefont {Wimmer}, \citenamefont {Kouwenhoven},\ and\ \citenamefont {Dvir}}]{Zatelli2023}%
  \BibitemOpen
  \bibfield  {author} {\bibinfo {author} {\bibfnamefont {F.}~\bibnamefont {Zatelli}}, \bibinfo {author} {\bibfnamefont {D.}~\bibnamefont {van Driel}}, \bibinfo {author} {\bibfnamefont {D.}~\bibnamefont {Xu}}, \bibinfo {author} {\bibfnamefont {G.}~\bibnamefont {Wang}}, \bibinfo {author} {\bibfnamefont {C.-X.}\ \bibnamefont {Liu}}, \bibinfo {author} {\bibfnamefont {A.}~\bibnamefont {Bordin}}, \bibinfo {author} {\bibfnamefont {B.}~\bibnamefont {Roovers}}, \bibinfo {author} {\bibfnamefont {G.~P.}\ \bibnamefont {Mazur}}, \bibinfo {author} {\bibfnamefont {N.}~\bibnamefont {van Loo}}, \bibinfo {author} {\bibfnamefont {J.~C.}\ \bibnamefont {Wolff}}, \bibinfo {author} {\bibfnamefont {A.~M.}\ \bibnamefont {Bozkurt}}, \bibinfo {author} {\bibfnamefont {G.}~\bibnamefont {Badawy}}, \bibinfo {author} {\bibfnamefont {S.}~\bibnamefont {Gazibegovic}}, \bibinfo {author} {\bibfnamefont {E.~P. A.~M.}\ \bibnamefont {Bakkers}}, \bibinfo {author} {\bibfnamefont {M.}~\bibnamefont {Wimmer}}, \bibinfo {author} {\bibfnamefont {L.~P.}\
  \bibnamefont {Kouwenhoven}},\ and\ \bibinfo {author} {\bibfnamefont {T.}~\bibnamefont {Dvir}},\ }\href {https://arxiv.org/abs/2311.03193v1} {\bibinfo {title} {Robust poor man's majorana zero modes using yu-shiba-rusinov states}} (\bibinfo {year} {2023})\BibitemShut {NoStop}%
\bibitem [{\citenamefont {Dvir}\ \emph {et~al.}(2023)\citenamefont {Dvir}, \citenamefont {Wang}, \citenamefont {van Loo}, \citenamefont {Liu}, \citenamefont {Mazur}, \citenamefont {Bordin}, \citenamefont {ten Haaf}, \citenamefont {Wang}, \citenamefont {van Driel}, \citenamefont {Zatelli}, \citenamefont {Li}, \citenamefont {Malinowski}, \citenamefont {Gazibegovic}, \citenamefont {Badawy}, \citenamefont {Bakkers}, \citenamefont {Wimmer},\ and\ \citenamefont {Kouwenhoven}}]{Dvir2023}%
  \BibitemOpen
  \bibfield  {author} {\bibinfo {author} {\bibfnamefont {T.}~\bibnamefont {Dvir}}, \bibinfo {author} {\bibfnamefont {G.}~\bibnamefont {Wang}}, \bibinfo {author} {\bibfnamefont {N.}~\bibnamefont {van Loo}}, \bibinfo {author} {\bibfnamefont {C.~X.}\ \bibnamefont {Liu}}, \bibinfo {author} {\bibfnamefont {G.~P.}\ \bibnamefont {Mazur}}, \bibinfo {author} {\bibfnamefont {A.}~\bibnamefont {Bordin}}, \bibinfo {author} {\bibfnamefont {S.~L.}\ \bibnamefont {ten Haaf}}, \bibinfo {author} {\bibfnamefont {J.~Y.}\ \bibnamefont {Wang}}, \bibinfo {author} {\bibfnamefont {D.}~\bibnamefont {van Driel}}, \bibinfo {author} {\bibfnamefont {F.}~\bibnamefont {Zatelli}}, \bibinfo {author} {\bibfnamefont {X.}~\bibnamefont {Li}}, \bibinfo {author} {\bibfnamefont {F.~K.}\ \bibnamefont {Malinowski}}, \bibinfo {author} {\bibfnamefont {S.}~\bibnamefont {Gazibegovic}}, \bibinfo {author} {\bibfnamefont {G.}~\bibnamefont {Badawy}}, \bibinfo {author} {\bibfnamefont {E.~P.}\ \bibnamefont {Bakkers}}, \bibinfo {author} {\bibfnamefont
  {M.}~\bibnamefont {Wimmer}},\ and\ \bibinfo {author} {\bibfnamefont {L.~P.}\ \bibnamefont {Kouwenhoven}},\ }\href {https://doi.org/10.1038/s41586-022-05585-1} {\bibfield  {journal} {\bibinfo  {journal} {Nature 2023 614:7948}\ }\textbf {\bibinfo {volume} {614}},\ \bibinfo {pages} {445} (\bibinfo {year} {2023})}\BibitemShut {NoStop}%
\bibitem [{\citenamefont {ten Haaf}\ \emph {et~al.}(2024)\citenamefont {ten Haaf}, \citenamefont {Wang}, \citenamefont {Bozkurt}, \citenamefont {Liu}, \citenamefont {Kulesh}, \citenamefont {Kim}, \citenamefont {Xiao}, \citenamefont {Thomas}, \citenamefont {Manfra}, \citenamefont {Dvir}, \citenamefont {Wimmer},\ and\ \citenamefont {Goswami}}]{Sebastiaan2024}%
  \BibitemOpen
  \bibfield  {author} {\bibinfo {author} {\bibfnamefont {S.~L.}\ \bibnamefont {ten Haaf}}, \bibinfo {author} {\bibfnamefont {Q.}~\bibnamefont {Wang}}, \bibinfo {author} {\bibfnamefont {A.~M.}\ \bibnamefont {Bozkurt}}, \bibinfo {author} {\bibfnamefont {C.~X.}\ \bibnamefont {Liu}}, \bibinfo {author} {\bibfnamefont {I.}~\bibnamefont {Kulesh}}, \bibinfo {author} {\bibfnamefont {P.}~\bibnamefont {Kim}}, \bibinfo {author} {\bibfnamefont {D.}~\bibnamefont {Xiao}}, \bibinfo {author} {\bibfnamefont {C.}~\bibnamefont {Thomas}}, \bibinfo {author} {\bibfnamefont {M.~J.}\ \bibnamefont {Manfra}}, \bibinfo {author} {\bibfnamefont {T.}~\bibnamefont {Dvir}}, \bibinfo {author} {\bibfnamefont {M.}~\bibnamefont {Wimmer}},\ and\ \bibinfo {author} {\bibfnamefont {S.}~\bibnamefont {Goswami}},\ }\href {https://doi.org/10.1038/s41586-024-07434-9} {\bibfield  {journal} {\bibinfo  {journal} {Nature 2024 630:8016}\ }\textbf {\bibinfo {volume} {630}},\ \bibinfo {pages} {329} (\bibinfo {year} {2024})}\BibitemShut {NoStop}%
\bibitem [{\citenamefont {Buitelaar}\ \emph {et~al.}(2002)\citenamefont {Buitelaar}, \citenamefont {Nussbaumer},\ and\ \citenamefont {Schönenberger}}]{Buitelaar2002}%
  \BibitemOpen
  \bibfield  {author} {\bibinfo {author} {\bibfnamefont {M.~R.}\ \bibnamefont {Buitelaar}}, \bibinfo {author} {\bibfnamefont {T.}~\bibnamefont {Nussbaumer}},\ and\ \bibinfo {author} {\bibfnamefont {C.}~\bibnamefont {Schönenberger}},\ }\href {https://doi.org/10.1103/PHYSREVLETT.89.256801/FIGURES/4/MEDIUM} {\bibfield  {journal} {\bibinfo  {journal} {Physical Review Letters}\ }\textbf {\bibinfo {volume} {89}},\ \bibinfo {pages} {256801} (\bibinfo {year} {2002})}\BibitemShut {NoStop}%
\bibitem [{\citenamefont {Pillet}\ \emph {et~al.}(2010)\citenamefont {Pillet}, \citenamefont {Quay}, \citenamefont {Morfin}, \citenamefont {Bena}, \citenamefont {Yeyati},\ and\ \citenamefont {Joyez}}]{Pillet2010}%
  \BibitemOpen
  \bibfield  {author} {\bibinfo {author} {\bibfnamefont {J.~D.}\ \bibnamefont {Pillet}}, \bibinfo {author} {\bibfnamefont {C.~H.}\ \bibnamefont {Quay}}, \bibinfo {author} {\bibfnamefont {P.}~\bibnamefont {Morfin}}, \bibinfo {author} {\bibfnamefont {C.}~\bibnamefont {Bena}}, \bibinfo {author} {\bibfnamefont {A.~L.}\ \bibnamefont {Yeyati}},\ and\ \bibinfo {author} {\bibfnamefont {P.}~\bibnamefont {Joyez}},\ }\href {https://doi.org/10.1038/nphys1811} {\bibfield  {journal} {\bibinfo  {journal} {Nature Physics 2010 6:12}\ }\textbf {\bibinfo {volume} {6}},\ \bibinfo {pages} {965} (\bibinfo {year} {2010})}\BibitemShut {NoStop}%
\bibitem [{\citenamefont {Lee}\ \emph {et~al.}(2013)\citenamefont {Lee}, \citenamefont {Jiang}, \citenamefont {Houzet}, \citenamefont {Aguado}, \citenamefont {Lieber},\ and\ \citenamefont {Franceschi}}]{Lee2013}%
  \BibitemOpen
  \bibfield  {author} {\bibinfo {author} {\bibfnamefont {E.~J.}\ \bibnamefont {Lee}}, \bibinfo {author} {\bibfnamefont {X.}~\bibnamefont {Jiang}}, \bibinfo {author} {\bibfnamefont {M.}~\bibnamefont {Houzet}}, \bibinfo {author} {\bibfnamefont {R.}~\bibnamefont {Aguado}}, \bibinfo {author} {\bibfnamefont {C.~M.}\ \bibnamefont {Lieber}},\ and\ \bibinfo {author} {\bibfnamefont {S.~D.}\ \bibnamefont {Franceschi}},\ }\href {https://doi.org/10.1038/nnano.2013.267} {\bibfield  {journal} {\bibinfo  {journal} {Nature Nanotechnology 2013 9:1}\ }\textbf {\bibinfo {volume} {9}},\ \bibinfo {pages} {79} (\bibinfo {year} {2013})}\BibitemShut {NoStop}%
\bibitem [{\citenamefont {Jellinggaard}\ \emph {et~al.}(2016)\citenamefont {Jellinggaard}, \citenamefont {Grove-Rasmussen}, \citenamefont {Madsen},\ and\ \citenamefont {Nygård}}]{Jellinggaard2016}%
  \BibitemOpen
  \bibfield  {author} {\bibinfo {author} {\bibfnamefont {A.}~\bibnamefont {Jellinggaard}}, \bibinfo {author} {\bibfnamefont {K.}~\bibnamefont {Grove-Rasmussen}}, \bibinfo {author} {\bibfnamefont {M.~H.}\ \bibnamefont {Madsen}},\ and\ \bibinfo {author} {\bibfnamefont {J.}~\bibnamefont {Nygård}},\ }\href {https://doi.org/10.1103/PHYSREVB.94.064520/FIGURES/10/MEDIUM} {\bibfield  {journal} {\bibinfo  {journal} {Physical Review B}\ }\textbf {\bibinfo {volume} {94}},\ \bibinfo {pages} {064520} (\bibinfo {year} {2016})}\BibitemShut {NoStop}%
\bibitem [{\citenamefont {Bretheau}\ \emph {et~al.}(2017)\citenamefont {Bretheau}, \citenamefont {Wang}, \citenamefont {Pisoni}, \citenamefont {Watanabe}, \citenamefont {Taniguchi},\ and\ \citenamefont {Jarillo-Herrero}}]{Bretheau2017}%
  \BibitemOpen
  \bibfield  {author} {\bibinfo {author} {\bibfnamefont {L.}~\bibnamefont {Bretheau}}, \bibinfo {author} {\bibfnamefont {J.~I.}\ \bibnamefont {Wang}}, \bibinfo {author} {\bibfnamefont {R.}~\bibnamefont {Pisoni}}, \bibinfo {author} {\bibfnamefont {K.}~\bibnamefont {Watanabe}}, \bibinfo {author} {\bibfnamefont {T.}~\bibnamefont {Taniguchi}},\ and\ \bibinfo {author} {\bibfnamefont {P.}~\bibnamefont {Jarillo-Herrero}},\ }\href {https://doi.org/10.1038/nphys4110} {\bibfield  {journal} {\bibinfo  {journal} {Nature Physics 2017 13:8}\ }\textbf {\bibinfo {volume} {13}},\ \bibinfo {pages} {756} (\bibinfo {year} {2017})}\BibitemShut {NoStop}%
\bibitem [{\citenamefont {Wang}\ \emph {et~al.}(2018)\citenamefont {Wang}, \citenamefont {Bretheau}, \citenamefont {Rodan-Legrain}, \citenamefont {Pisoni}, \citenamefont {Watanabe}, \citenamefont {Taniguchi},\ and\ \citenamefont {Jarillo-Herrero}}]{Wang2018a}%
  \BibitemOpen
  \bibfield  {author} {\bibinfo {author} {\bibfnamefont {J.~I.}\ \bibnamefont {Wang}}, \bibinfo {author} {\bibfnamefont {L.}~\bibnamefont {Bretheau}}, \bibinfo {author} {\bibfnamefont {D.}~\bibnamefont {Rodan-Legrain}}, \bibinfo {author} {\bibfnamefont {R.}~\bibnamefont {Pisoni}}, \bibinfo {author} {\bibfnamefont {K.}~\bibnamefont {Watanabe}}, \bibinfo {author} {\bibfnamefont {T.}~\bibnamefont {Taniguchi}},\ and\ \bibinfo {author} {\bibfnamefont {P.}~\bibnamefont {Jarillo-Herrero}},\ }\href {https://doi.org/10.1103/PhysRevB.98.121411} {\bibfield  {journal} {\bibinfo  {journal} {Physical Review B}\ }\textbf {\bibinfo {volume} {98}},\ \bibinfo {pages} {1} (\bibinfo {year} {2018})}\BibitemShut {NoStop}%
\bibitem [{\citenamefont {Prada}\ \emph {et~al.}(2020)\citenamefont {Prada}, \citenamefont {San-Jose}, \citenamefont {de~Moor}, \citenamefont {Geresdi}, \citenamefont {Lee}, \citenamefont {Klinovaja}, \citenamefont {Loss}, \citenamefont {Nygård}, \citenamefont {Aguado},\ and\ \citenamefont {Kouwenhoven}}]{Prada2020}%
  \BibitemOpen
  \bibfield  {author} {\bibinfo {author} {\bibfnamefont {E.}~\bibnamefont {Prada}}, \bibinfo {author} {\bibfnamefont {P.}~\bibnamefont {San-Jose}}, \bibinfo {author} {\bibfnamefont {M.~W.}\ \bibnamefont {de~Moor}}, \bibinfo {author} {\bibfnamefont {A.}~\bibnamefont {Geresdi}}, \bibinfo {author} {\bibfnamefont {E.~J.}\ \bibnamefont {Lee}}, \bibinfo {author} {\bibfnamefont {J.}~\bibnamefont {Klinovaja}}, \bibinfo {author} {\bibfnamefont {D.}~\bibnamefont {Loss}}, \bibinfo {author} {\bibfnamefont {J.}~\bibnamefont {Nygård}}, \bibinfo {author} {\bibfnamefont {R.}~\bibnamefont {Aguado}},\ and\ \bibinfo {author} {\bibfnamefont {L.~P.}\ \bibnamefont {Kouwenhoven}},\ }\href {https://doi.org/10.1038/s42254-020-0228-y} {\bibfield  {journal} {\bibinfo  {journal} {Nature Reviews Physics 2020 2:10}\ }\textbf {\bibinfo {volume} {2}},\ \bibinfo {pages} {575} (\bibinfo {year} {2020})}\BibitemShut {NoStop}%
\bibitem [{\citenamefont {van Driel}\ \emph {et~al.}(2023)\citenamefont {van Driel}, \citenamefont {Wang}, \citenamefont {Bordin}, \citenamefont {van Loo}, \citenamefont {Zatelli}, \citenamefont {Mazur}, \citenamefont {Xu}, \citenamefont {Gazibegovic}, \citenamefont {Badawy}, \citenamefont {Bakkers}, \citenamefont {Kouwenhoven},\ and\ \citenamefont {Dvir}}]{vanDriel2023}%
  \BibitemOpen
  \bibfield  {author} {\bibinfo {author} {\bibfnamefont {D.}~\bibnamefont {van Driel}}, \bibinfo {author} {\bibfnamefont {G.}~\bibnamefont {Wang}}, \bibinfo {author} {\bibfnamefont {A.}~\bibnamefont {Bordin}}, \bibinfo {author} {\bibfnamefont {N.}~\bibnamefont {van Loo}}, \bibinfo {author} {\bibfnamefont {F.}~\bibnamefont {Zatelli}}, \bibinfo {author} {\bibfnamefont {G.~P.}\ \bibnamefont {Mazur}}, \bibinfo {author} {\bibfnamefont {D.}~\bibnamefont {Xu}}, \bibinfo {author} {\bibfnamefont {S.}~\bibnamefont {Gazibegovic}}, \bibinfo {author} {\bibfnamefont {G.}~\bibnamefont {Badawy}}, \bibinfo {author} {\bibfnamefont {E.~P.}\ \bibnamefont {Bakkers}}, \bibinfo {author} {\bibfnamefont {L.~P.}\ \bibnamefont {Kouwenhoven}},\ and\ \bibinfo {author} {\bibfnamefont {T.}~\bibnamefont {Dvir}},\ }\href {https://doi.org/10.1038/s41467-023-42026-7} {\bibfield  {journal} {\bibinfo  {journal} {Nature Communications 2023 14:1}\ }\textbf {\bibinfo {volume} {14}},\ \bibinfo {pages} {1} (\bibinfo {year} {2023})}\BibitemShut
  {NoStop}%
\bibitem [{\citenamefont {Bretheau}\ \emph {et~al.}(2013)\citenamefont {Bretheau}, \citenamefont {\c{C} Ö~Girit}, \citenamefont {Pothier}, \citenamefont {Esteve},\ and\ \citenamefont {Urbina}}]{Bretheau2013}%
  \BibitemOpen
  \bibfield  {author} {\bibinfo {author} {\bibfnamefont {L.}~\bibnamefont {Bretheau}}, \bibinfo {author} {\bibnamefont {\c{C} Ö~Girit}}, \bibinfo {author} {\bibfnamefont {H.}~\bibnamefont {Pothier}}, \bibinfo {author} {\bibfnamefont {D.}~\bibnamefont {Esteve}},\ and\ \bibinfo {author} {\bibfnamefont {C.}~\bibnamefont {Urbina}},\ }\href {https://doi.org/10.1038/nature12315} {\bibfield  {journal} {\bibinfo  {journal} {Nature 2013 499:7458}\ }\textbf {\bibinfo {volume} {499}},\ \bibinfo {pages} {312} (\bibinfo {year} {2013})}\BibitemShut {NoStop}%
\bibitem [{\citenamefont {Woerkom}\ \emph {et~al.}(2017)\citenamefont {Woerkom}, \citenamefont {Proutski}, \citenamefont {Heck}, \citenamefont {Bouman}, \citenamefont {Väyrynen}, \citenamefont {Glazman}, \citenamefont {Krogstrup}, \citenamefont {Nygård}, \citenamefont {Kouwenhoven},\ and\ \citenamefont {Geresdi}}]{Woerkom2017}%
  \BibitemOpen
  \bibfield  {author} {\bibinfo {author} {\bibfnamefont {D.~J.~V.}\ \bibnamefont {Woerkom}}, \bibinfo {author} {\bibfnamefont {A.}~\bibnamefont {Proutski}}, \bibinfo {author} {\bibfnamefont {B.~V.}\ \bibnamefont {Heck}}, \bibinfo {author} {\bibfnamefont {D.}~\bibnamefont {Bouman}}, \bibinfo {author} {\bibfnamefont {J.~I.}\ \bibnamefont {Väyrynen}}, \bibinfo {author} {\bibfnamefont {L.~I.}\ \bibnamefont {Glazman}}, \bibinfo {author} {\bibfnamefont {P.}~\bibnamefont {Krogstrup}}, \bibinfo {author} {\bibfnamefont {J.}~\bibnamefont {Nygård}}, \bibinfo {author} {\bibfnamefont {L.~P.}\ \bibnamefont {Kouwenhoven}},\ and\ \bibinfo {author} {\bibfnamefont {A.}~\bibnamefont {Geresdi}},\ }\href {https://doi.org/10.1038/nphys4150} {\bibfield  {journal} {\bibinfo  {journal} {Nature Physics 2017 13:9}\ }\textbf {\bibinfo {volume} {13}},\ \bibinfo {pages} {876} (\bibinfo {year} {2017})}\BibitemShut {NoStop}%
\bibitem [{\citenamefont {Dassonneville}\ \emph {et~al.}(2018)\citenamefont {Dassonneville}, \citenamefont {Murani}, \citenamefont {Ferrier}, \citenamefont {Gu\'eron},\ and\ \citenamefont {Bouchiat}}]{Dassonneville2018}%
  \BibitemOpen
  \bibfield  {author} {\bibinfo {author} {\bibfnamefont {B.}~\bibnamefont {Dassonneville}}, \bibinfo {author} {\bibfnamefont {A.}~\bibnamefont {Murani}}, \bibinfo {author} {\bibfnamefont {M.}~\bibnamefont {Ferrier}}, \bibinfo {author} {\bibfnamefont {S.}~\bibnamefont {Gu\'eron}},\ and\ \bibinfo {author} {\bibfnamefont {H.}~\bibnamefont {Bouchiat}},\ }\href {https://doi.org/10.1103/PhysRevB.97.184505} {\bibfield  {journal} {\bibinfo  {journal} {Phys. Rev. B}\ }\textbf {\bibinfo {volume} {97}},\ \bibinfo {pages} {184505} (\bibinfo {year} {2018})}\BibitemShut {NoStop}%
\bibitem [{\citenamefont {Tosi}\ \emph {et~al.}(2019)\citenamefont {Tosi}, \citenamefont {Metzger}, \citenamefont {Goffman}, \citenamefont {Urbina}, \citenamefont {Pothier}, \citenamefont {Park}, \citenamefont {Yeyati}, \citenamefont {Nyg\aa{}rd},\ and\ \citenamefont {Krogstrup}}]{Tosi2019}%
  \BibitemOpen
  \bibfield  {author} {\bibinfo {author} {\bibfnamefont {L.}~\bibnamefont {Tosi}}, \bibinfo {author} {\bibfnamefont {C.}~\bibnamefont {Metzger}}, \bibinfo {author} {\bibfnamefont {M.~F.}\ \bibnamefont {Goffman}}, \bibinfo {author} {\bibfnamefont {C.}~\bibnamefont {Urbina}}, \bibinfo {author} {\bibfnamefont {H.}~\bibnamefont {Pothier}}, \bibinfo {author} {\bibfnamefont {S.}~\bibnamefont {Park}}, \bibinfo {author} {\bibfnamefont {A.~L.}\ \bibnamefont {Yeyati}}, \bibinfo {author} {\bibfnamefont {J.}~\bibnamefont {Nyg\aa{}rd}},\ and\ \bibinfo {author} {\bibfnamefont {P.}~\bibnamefont {Krogstrup}},\ }\href {https://doi.org/10.1103/PhysRevX.9.011010} {\bibfield  {journal} {\bibinfo  {journal} {Phys. Rev. X}\ }\textbf {\bibinfo {volume} {9}},\ \bibinfo {pages} {011010} (\bibinfo {year} {2019})}\BibitemShut {NoStop}%
\bibitem [{\citenamefont {Chidambaram}\ \emph {et~al.}(2022)\citenamefont {Chidambaram}, \citenamefont {Kringh\o{}j}, \citenamefont {Casparis}, \citenamefont {Kuemmeth}, \citenamefont {Wang}, \citenamefont {Thomas}, \citenamefont {Gronin}, \citenamefont {Gardner}, \citenamefont {Cui}, \citenamefont {Liu}, \citenamefont {Moors}, \citenamefont {Manfra}, \citenamefont {Petersson},\ and\ \citenamefont {Connolly}}]{Chidambaram2022}%
  \BibitemOpen
  \bibfield  {author} {\bibinfo {author} {\bibfnamefont {V.}~\bibnamefont {Chidambaram}}, \bibinfo {author} {\bibfnamefont {A.}~\bibnamefont {Kringh\o{}j}}, \bibinfo {author} {\bibfnamefont {L.}~\bibnamefont {Casparis}}, \bibinfo {author} {\bibfnamefont {F.}~\bibnamefont {Kuemmeth}}, \bibinfo {author} {\bibfnamefont {T.}~\bibnamefont {Wang}}, \bibinfo {author} {\bibfnamefont {C.}~\bibnamefont {Thomas}}, \bibinfo {author} {\bibfnamefont {S.}~\bibnamefont {Gronin}}, \bibinfo {author} {\bibfnamefont {G.~C.}\ \bibnamefont {Gardner}}, \bibinfo {author} {\bibfnamefont {Z.}~\bibnamefont {Cui}}, \bibinfo {author} {\bibfnamefont {C.}~\bibnamefont {Liu}}, \bibinfo {author} {\bibfnamefont {K.}~\bibnamefont {Moors}}, \bibinfo {author} {\bibfnamefont {M.~J.}\ \bibnamefont {Manfra}}, \bibinfo {author} {\bibfnamefont {K.~D.}\ \bibnamefont {Petersson}},\ and\ \bibinfo {author} {\bibfnamefont {M.~R.}\ \bibnamefont {Connolly}},\ }\href {https://doi.org/10.1103/PhysRevResearch.4.023170} {\bibfield  {journal} {\bibinfo
  {journal} {Phys. Rev. Res.}\ }\textbf {\bibinfo {volume} {4}},\ \bibinfo {pages} {023170} (\bibinfo {year} {2022})}\BibitemShut {NoStop}%
\bibitem [{\citenamefont {Hinderling}\ \emph {et~al.}(2023)\citenamefont {Hinderling}, \citenamefont {Sabonis}, \citenamefont {Paredes}, \citenamefont {Haxell}, \citenamefont {Coraiola}, \citenamefont {ten Kate}, \citenamefont {Cheah}, \citenamefont {Krizek}, \citenamefont {Schott}, \citenamefont {Wegscheider},\ and\ \citenamefont {Nichele}}]{Hinderling2023}%
  \BibitemOpen
  \bibfield  {author} {\bibinfo {author} {\bibfnamefont {M.}~\bibnamefont {Hinderling}}, \bibinfo {author} {\bibfnamefont {D.}~\bibnamefont {Sabonis}}, \bibinfo {author} {\bibfnamefont {S.}~\bibnamefont {Paredes}}, \bibinfo {author} {\bibfnamefont {D.}~\bibnamefont {Haxell}}, \bibinfo {author} {\bibfnamefont {M.}~\bibnamefont {Coraiola}}, \bibinfo {author} {\bibfnamefont {S.}~\bibnamefont {ten Kate}}, \bibinfo {author} {\bibfnamefont {E.}~\bibnamefont {Cheah}}, \bibinfo {author} {\bibfnamefont {F.}~\bibnamefont {Krizek}}, \bibinfo {author} {\bibfnamefont {R.}~\bibnamefont {Schott}}, \bibinfo {author} {\bibfnamefont {W.}~\bibnamefont {Wegscheider}},\ and\ \bibinfo {author} {\bibfnamefont {F.}~\bibnamefont {Nichele}},\ }\href {https://doi.org/10.1103/PhysRevApplied.19.054026} {\bibfield  {journal} {\bibinfo  {journal} {Phys. Rev. Appl.}\ }\textbf {\bibinfo {volume} {19}},\ \bibinfo {pages} {054026} (\bibinfo {year} {2023})}\BibitemShut {NoStop}%
\bibitem [{\citenamefont {Szombati}\ \emph {et~al.}(2016)\citenamefont {Szombati}, \citenamefont {Nadj-Perge}, \citenamefont {Car}, \citenamefont {Plissard}, \citenamefont {Bakkers},\ and\ \citenamefont {Kouwenhoven}}]{Szombati2016}%
  \BibitemOpen
  \bibfield  {author} {\bibinfo {author} {\bibfnamefont {D.~B.}\ \bibnamefont {Szombati}}, \bibinfo {author} {\bibfnamefont {S.}~\bibnamefont {Nadj-Perge}}, \bibinfo {author} {\bibfnamefont {D.}~\bibnamefont {Car}}, \bibinfo {author} {\bibfnamefont {S.~R.}\ \bibnamefont {Plissard}}, \bibinfo {author} {\bibfnamefont {E.~P.}\ \bibnamefont {Bakkers}},\ and\ \bibinfo {author} {\bibfnamefont {L.~P.}\ \bibnamefont {Kouwenhoven}},\ }\href {https://doi.org/10.1038/nphys3742} {\bibfield  {journal} {\bibinfo  {journal} {Nature Physics 2016 12:6}\ }\textbf {\bibinfo {volume} {12}},\ \bibinfo {pages} {568} (\bibinfo {year} {2016})}\BibitemShut {NoStop}%
\bibitem [{\citenamefont {Nanda}\ \emph {et~al.}(2017)\citenamefont {Nanda}, \citenamefont {Aguilera-Servin}, \citenamefont {Rakyta}, \citenamefont {Kormányos}, \citenamefont {Kleiner}, \citenamefont {Koelle}, \citenamefont {Watanabe}, \citenamefont {Taniguchi}, \citenamefont {Vandersypen},\ and\ \citenamefont {Goswami}}]{Nanda2017}%
  \BibitemOpen
  \bibfield  {author} {\bibinfo {author} {\bibfnamefont {G.}~\bibnamefont {Nanda}}, \bibinfo {author} {\bibfnamefont {J.~L.}\ \bibnamefont {Aguilera-Servin}}, \bibinfo {author} {\bibfnamefont {P.}~\bibnamefont {Rakyta}}, \bibinfo {author} {\bibfnamefont {A.}~\bibnamefont {Kormányos}}, \bibinfo {author} {\bibfnamefont {R.}~\bibnamefont {Kleiner}}, \bibinfo {author} {\bibfnamefont {D.}~\bibnamefont {Koelle}}, \bibinfo {author} {\bibfnamefont {K.}~\bibnamefont {Watanabe}}, \bibinfo {author} {\bibfnamefont {T.}~\bibnamefont {Taniguchi}}, \bibinfo {author} {\bibfnamefont {L.~M.~K.}\ \bibnamefont {Vandersypen}},\ and\ \bibinfo {author} {\bibfnamefont {S.}~\bibnamefont {Goswami}},\ }\href {https://doi.org/10.1021/acs.nanolett.7b00097} {\bibfield  {journal} {\bibinfo  {journal} {Nano Letters}\ }\textbf {\bibinfo {volume} {17}},\ \bibinfo {pages} {3396} (\bibinfo {year} {2017})},\ \bibinfo {note} {pMID: 28474892}\BibitemShut {NoStop}%
\bibitem [{\citenamefont {Spanton}\ \emph {et~al.}(2017)\citenamefont {Spanton}, \citenamefont {Deng}, \citenamefont {Vaitiekėnas}, \citenamefont {Krogstrup}, \citenamefont {Nyg{\aa}rd}, \citenamefont {Marcus},\ and\ \citenamefont {Moler}}]{Spanton2017open}%
  \BibitemOpen
  \bibfield  {author} {\bibinfo {author} {\bibfnamefont {E.~M.}\ \bibnamefont {Spanton}}, \bibinfo {author} {\bibfnamefont {M.}~\bibnamefont {Deng}}, \bibinfo {author} {\bibfnamefont {S.}~\bibnamefont {Vaitiekėnas}}, \bibinfo {author} {\bibfnamefont {P.}~\bibnamefont {Krogstrup}}, \bibinfo {author} {\bibfnamefont {J.}~\bibnamefont {Nyg{\aa}rd}}, \bibinfo {author} {\bibfnamefont {C.~M.}\ \bibnamefont {Marcus}},\ and\ \bibinfo {author} {\bibfnamefont {K.~A.}\ \bibnamefont {Moler}},\ }\href {https://doi.org/10.1038/nphys4224} {\bibfield  {journal} {\bibinfo  {journal} {Nat. Phys.}\ }\textbf {\bibinfo {volume} {13}},\ \bibinfo {pages} {1177} (\bibinfo {year} {2017})}\BibitemShut {NoStop}%
\bibitem [{\citenamefont {Della~Rocca}\ \emph {et~al.}(2007)\citenamefont {Della~Rocca}, \citenamefont {Chauvin}, \citenamefont {Huard}, \citenamefont {Pothier}, \citenamefont {Esteve},\ and\ \citenamefont {Urbina}}]{PhysRevLett.99.127005}%
  \BibitemOpen
  \bibfield  {author} {\bibinfo {author} {\bibfnamefont {M.~L.}\ \bibnamefont {Della~Rocca}}, \bibinfo {author} {\bibfnamefont {M.}~\bibnamefont {Chauvin}}, \bibinfo {author} {\bibfnamefont {B.}~\bibnamefont {Huard}}, \bibinfo {author} {\bibfnamefont {H.}~\bibnamefont {Pothier}}, \bibinfo {author} {\bibfnamefont {D.}~\bibnamefont {Esteve}},\ and\ \bibinfo {author} {\bibfnamefont {C.}~\bibnamefont {Urbina}},\ }\href {https://doi.org/10.1103/PhysRevLett.99.127005} {\bibfield  {journal} {\bibinfo  {journal} {Phys. Rev. Lett.}\ }\textbf {\bibinfo {volume} {99}},\ \bibinfo {pages} {127005} (\bibinfo {year} {2007})}\BibitemShut {NoStop}%
\bibitem [{\citenamefont {Indolese}\ \emph {et~al.}(2020)\citenamefont {Indolese}, \citenamefont {Karnatak}, \citenamefont {Kononov}, \citenamefont {Delagrange}, \citenamefont {Haller}, \citenamefont {Wang}, \citenamefont {Makk}, \citenamefont {Watanabe}, \citenamefont {Taniguchi},\ and\ \citenamefont {Schönenberger}}]{Indolese2020}%
  \BibitemOpen
  \bibfield  {author} {\bibinfo {author} {\bibfnamefont {D.~I.}\ \bibnamefont {Indolese}}, \bibinfo {author} {\bibfnamefont {P.}~\bibnamefont {Karnatak}}, \bibinfo {author} {\bibfnamefont {A.}~\bibnamefont {Kononov}}, \bibinfo {author} {\bibfnamefont {R.}~\bibnamefont {Delagrange}}, \bibinfo {author} {\bibfnamefont {R.}~\bibnamefont {Haller}}, \bibinfo {author} {\bibfnamefont {L.}~\bibnamefont {Wang}}, \bibinfo {author} {\bibfnamefont {P.}~\bibnamefont {Makk}}, \bibinfo {author} {\bibfnamefont {K.}~\bibnamefont {Watanabe}}, \bibinfo {author} {\bibfnamefont {T.}~\bibnamefont {Taniguchi}},\ and\ \bibinfo {author} {\bibfnamefont {C.}~\bibnamefont {Schönenberger}},\ }\href {https://doi.org/10.1021/ACS.NANOLETT.0C02412} {\bibfield  {journal} {\bibinfo  {journal} {Nano Letters}\ }\textbf {\bibinfo {volume} {20}},\ \bibinfo {pages} {7129} (\bibinfo {year} {2020})}\BibitemShut {NoStop}%
\bibitem [{\citenamefont {Haller}\ \emph {et~al.}(2022)\citenamefont {Haller}, \citenamefont {F\"ul\"op}, \citenamefont {Indolese}, \citenamefont {Ridderbos}, \citenamefont {Kraft}, \citenamefont {Cheung}, \citenamefont {Ungerer}, \citenamefont {Watanabe}, \citenamefont {Taniguchi}, \citenamefont {Beckmann}, \citenamefont {Danneau}, \citenamefont {Virtanen},\ and\ \citenamefont {Sch\"onenberger}}]{Haller2022}%
  \BibitemOpen
  \bibfield  {author} {\bibinfo {author} {\bibfnamefont {R.}~\bibnamefont {Haller}}, \bibinfo {author} {\bibfnamefont {G.}~\bibnamefont {F\"ul\"op}}, \bibinfo {author} {\bibfnamefont {D.}~\bibnamefont {Indolese}}, \bibinfo {author} {\bibfnamefont {J.}~\bibnamefont {Ridderbos}}, \bibinfo {author} {\bibfnamefont {R.}~\bibnamefont {Kraft}}, \bibinfo {author} {\bibfnamefont {L.~Y.}\ \bibnamefont {Cheung}}, \bibinfo {author} {\bibfnamefont {J.~H.}\ \bibnamefont {Ungerer}}, \bibinfo {author} {\bibfnamefont {K.}~\bibnamefont {Watanabe}}, \bibinfo {author} {\bibfnamefont {T.}~\bibnamefont {Taniguchi}}, \bibinfo {author} {\bibfnamefont {D.}~\bibnamefont {Beckmann}}, \bibinfo {author} {\bibfnamefont {R.}~\bibnamefont {Danneau}}, \bibinfo {author} {\bibfnamefont {P.}~\bibnamefont {Virtanen}},\ and\ \bibinfo {author} {\bibfnamefont {C.}~\bibnamefont {Sch\"onenberger}},\ }\href {https://doi.org/10.1103/PhysRevResearch.4.013198} {\bibfield  {journal} {\bibinfo  {journal} {Phys. Rev. Res.}\ }\textbf {\bibinfo {volume}
  {4}},\ \bibinfo {pages} {013198} (\bibinfo {year} {2022})}\BibitemShut {NoStop}%
\bibitem [{\citenamefont {Haxell}\ \emph {et~al.}(2023)\citenamefont {Haxell}, \citenamefont {Coraiola}, \citenamefont {Sabonis}, \citenamefont {Hinderling}, \citenamefont {ten Kate}, \citenamefont {Cheah}, \citenamefont {Krizek}, \citenamefont {Schott}, \citenamefont {Wegscheider},\ and\ \citenamefont {Nichele}}]{Haxell2023}%
  \BibitemOpen
  \bibfield  {author} {\bibinfo {author} {\bibfnamefont {D.~Z.}\ \bibnamefont {Haxell}}, \bibinfo {author} {\bibfnamefont {M.}~\bibnamefont {Coraiola}}, \bibinfo {author} {\bibfnamefont {D.}~\bibnamefont {Sabonis}}, \bibinfo {author} {\bibfnamefont {M.}~\bibnamefont {Hinderling}}, \bibinfo {author} {\bibfnamefont {S.~C.}\ \bibnamefont {ten Kate}}, \bibinfo {author} {\bibfnamefont {E.}~\bibnamefont {Cheah}}, \bibinfo {author} {\bibfnamefont {F.}~\bibnamefont {Krizek}}, \bibinfo {author} {\bibfnamefont {R.}~\bibnamefont {Schott}}, \bibinfo {author} {\bibfnamefont {W.}~\bibnamefont {Wegscheider}},\ and\ \bibinfo {author} {\bibfnamefont {F.}~\bibnamefont {Nichele}},\ }\href {https://doi.org/10.1021/acsnano.3c04957} {\bibfield  {journal} {\bibinfo  {journal} {ACS Nano}\ }\textbf {\bibinfo {volume} {17}},\ \bibinfo {pages} {18139} (\bibinfo {year} {2023})}\BibitemShut {NoStop}%
\bibitem [{\citenamefont {Strickland}\ \emph {et~al.}(2022)\citenamefont {Strickland}, \citenamefont {Elfeky}, \citenamefont {Yuan}, \citenamefont {Schiela}, \citenamefont {Yu}, \citenamefont {Langone}, \citenamefont {Vavilov}, \citenamefont {Manucharyan},\ and\ \citenamefont {Shabani}}]{Strickland2022}%
  \BibitemOpen
  \bibfield  {author} {\bibinfo {author} {\bibfnamefont {W.~M.}\ \bibnamefont {Strickland}}, \bibinfo {author} {\bibfnamefont {B.~H.}\ \bibnamefont {Elfeky}}, \bibinfo {author} {\bibfnamefont {J.~O.}\ \bibnamefont {Yuan}}, \bibinfo {author} {\bibfnamefont {W.~F.}\ \bibnamefont {Schiela}}, \bibinfo {author} {\bibfnamefont {P.}~\bibnamefont {Yu}}, \bibinfo {author} {\bibfnamefont {D.}~\bibnamefont {Langone}}, \bibinfo {author} {\bibfnamefont {M.~G.}\ \bibnamefont {Vavilov}}, \bibinfo {author} {\bibfnamefont {V.~E.}\ \bibnamefont {Manucharyan}},\ and\ \bibinfo {author} {\bibfnamefont {J.}~\bibnamefont {Shabani}},\ }\bibfield  {journal} {\bibinfo  {journal} {Physical Review Applied}\ }\textbf {\bibinfo {volume} {19}},\ \href {https://doi.org/10.1103/PhysRevApplied.19.034021} {10.1103/PhysRevApplied.19.034021} (\bibinfo {year} {2022})\BibitemShut {NoStop}%
\bibitem [{\citenamefont {Elfeky}\ \emph {et~al.}(2023)\citenamefont {Elfeky}, \citenamefont {Strickland}, \citenamefont {Lee}, \citenamefont {Farmer}, \citenamefont {Shanto}, \citenamefont {Zarassi}, \citenamefont {Langone}, \citenamefont {Vavilov}, \citenamefont {Levenson-Falk},\ and\ \citenamefont {Shabani}}]{Elfeky2023}%
  \BibitemOpen
  \bibfield  {author} {\bibinfo {author} {\bibfnamefont {B.~H.}\ \bibnamefont {Elfeky}}, \bibinfo {author} {\bibfnamefont {W.~M.}\ \bibnamefont {Strickland}}, \bibinfo {author} {\bibfnamefont {J.}~\bibnamefont {Lee}}, \bibinfo {author} {\bibfnamefont {J.~T.}\ \bibnamefont {Farmer}}, \bibinfo {author} {\bibfnamefont {S.}~\bibnamefont {Shanto}}, \bibinfo {author} {\bibfnamefont {A.}~\bibnamefont {Zarassi}}, \bibinfo {author} {\bibfnamefont {D.}~\bibnamefont {Langone}}, \bibinfo {author} {\bibfnamefont {M.~G.}\ \bibnamefont {Vavilov}}, \bibinfo {author} {\bibfnamefont {E.~M.}\ \bibnamefont {Levenson-Falk}},\ and\ \bibinfo {author} {\bibfnamefont {J.}~\bibnamefont {Shabani}},\ }\bibfield  {journal} {\bibinfo  {journal} {PRX Quantum}\ }\textbf {\bibinfo {volume} {4}},\ \href {https://doi.org/10.1103/PRXQuantum.4.030339} {10.1103/PRXQuantum.4.030339} (\bibinfo {year} {2023})\BibitemShut {NoStop}%
\bibitem [{\citenamefont {Phan}\ \emph {et~al.}(2022)\citenamefont {Phan}, \citenamefont {Senior}, \citenamefont {Ghazaryan}, \citenamefont {Hatefipour}, \citenamefont {Strickland}, \citenamefont {Shabani}, \citenamefont {Serbyn},\ and\ \citenamefont {Higginbotham}}]{Phan2022}%
  \BibitemOpen
  \bibfield  {author} {\bibinfo {author} {\bibfnamefont {D.}~\bibnamefont {Phan}}, \bibinfo {author} {\bibfnamefont {J.}~\bibnamefont {Senior}}, \bibinfo {author} {\bibfnamefont {A.}~\bibnamefont {Ghazaryan}}, \bibinfo {author} {\bibfnamefont {M.}~\bibnamefont {Hatefipour}}, \bibinfo {author} {\bibfnamefont {W.~M.}\ \bibnamefont {Strickland}}, \bibinfo {author} {\bibfnamefont {J.}~\bibnamefont {Shabani}}, \bibinfo {author} {\bibfnamefont {M.}~\bibnamefont {Serbyn}},\ and\ \bibinfo {author} {\bibfnamefont {A.~P.}\ \bibnamefont {Higginbotham}},\ }\href {https://doi.org/10.1103/PhysRevLett.128.107701} {\bibfield  {journal} {\bibinfo  {journal} {Phys. Rev. Lett.}\ }\textbf {\bibinfo {volume} {128}},\ \bibinfo {pages} {107701} (\bibinfo {year} {2022})}\BibitemShut {NoStop}%
\bibitem [{\citenamefont {Phan}\ \emph {et~al.}(2023)\citenamefont {Phan}, \citenamefont {Falthansl-Scheinecker}, \citenamefont {Mishra}, \citenamefont {Strickland}, \citenamefont {Langone}, \citenamefont {Shabani},\ and\ \citenamefont {Higginbotham}}]{Phan2023}%
  \BibitemOpen
  \bibfield  {author} {\bibinfo {author} {\bibfnamefont {D.}~\bibnamefont {Phan}}, \bibinfo {author} {\bibfnamefont {P.}~\bibnamefont {Falthansl-Scheinecker}}, \bibinfo {author} {\bibfnamefont {U.}~\bibnamefont {Mishra}}, \bibinfo {author} {\bibfnamefont {W.}~\bibnamefont {Strickland}}, \bibinfo {author} {\bibfnamefont {D.}~\bibnamefont {Langone}}, \bibinfo {author} {\bibfnamefont {J.}~\bibnamefont {Shabani}},\ and\ \bibinfo {author} {\bibfnamefont {A.}~\bibnamefont {Higginbotham}},\ }\href {https://doi.org/10.1103/PhysRevApplied.19.064032} {\bibfield  {journal} {\bibinfo  {journal} {Phys. Rev. Appl.}\ }\textbf {\bibinfo {volume} {19}},\ \bibinfo {pages} {064032} (\bibinfo {year} {2023})}\BibitemShut {NoStop}%
\bibitem [{\citenamefont {Babkin}\ \emph {et~al.}(2024)\citenamefont {Babkin}, \citenamefont {Higginbotham},\ and\ \citenamefont {Serbyn}}]{Serafim2024}%
  \BibitemOpen
  \bibfield  {author} {\bibinfo {author} {\bibfnamefont {S.~S.}\ \bibnamefont {Babkin}}, \bibinfo {author} {\bibfnamefont {A.~P.}\ \bibnamefont {Higginbotham}},\ and\ \bibinfo {author} {\bibfnamefont {M.}~\bibnamefont {Serbyn}},\ }\href {https://doi.org/10.21468/SciPostPhys.16.5.115} {\bibfield  {journal} {\bibinfo  {journal} {SciPost Phys.}\ }\textbf {\bibinfo {volume} {16}},\ \bibinfo {pages} {115} (\bibinfo {year} {2024})}\BibitemShut {NoStop}%
\bibitem [{\citenamefont {Willsch}\ \emph {et~al.}(2024)\citenamefont {Willsch}, \citenamefont {Rieger}, \citenamefont {Winkel}, \citenamefont {Willsch}, \citenamefont {Dickel}, \citenamefont {Krause}, \citenamefont {Ando}, \citenamefont {Lescanne}, \citenamefont {Leghtas}, \citenamefont {Bronn}, \citenamefont {Deb}, \citenamefont {Lanes}, \citenamefont {Minev}, \citenamefont {Dennig}, \citenamefont {Geisert}, \citenamefont {Günzler}, \citenamefont {Ihssen}, \citenamefont {Paluch}, \citenamefont {Reisinger}, \citenamefont {Hanna}, \citenamefont {Bae}, \citenamefont {Schüffelgen}, \citenamefont {Grützmacher}, \citenamefont {Buimaga-Iarinca}, \citenamefont {Morari}, \citenamefont {Wernsdorfer}, \citenamefont {DiVincenzo}, \citenamefont {Michielsen}, \citenamefont {Catelani},\ and\ \citenamefont {Pop}}]{Willsch2024}%
  \BibitemOpen
  \bibfield  {author} {\bibinfo {author} {\bibfnamefont {D.}~\bibnamefont {Willsch}}, \bibinfo {author} {\bibfnamefont {D.}~\bibnamefont {Rieger}}, \bibinfo {author} {\bibfnamefont {P.}~\bibnamefont {Winkel}}, \bibinfo {author} {\bibfnamefont {M.}~\bibnamefont {Willsch}}, \bibinfo {author} {\bibfnamefont {C.}~\bibnamefont {Dickel}}, \bibinfo {author} {\bibfnamefont {J.}~\bibnamefont {Krause}}, \bibinfo {author} {\bibfnamefont {Y.}~\bibnamefont {Ando}}, \bibinfo {author} {\bibfnamefont {R.}~\bibnamefont {Lescanne}}, \bibinfo {author} {\bibfnamefont {Z.}~\bibnamefont {Leghtas}}, \bibinfo {author} {\bibfnamefont {N.~T.}\ \bibnamefont {Bronn}}, \bibinfo {author} {\bibfnamefont {P.}~\bibnamefont {Deb}}, \bibinfo {author} {\bibfnamefont {O.}~\bibnamefont {Lanes}}, \bibinfo {author} {\bibfnamefont {Z.~K.}\ \bibnamefont {Minev}}, \bibinfo {author} {\bibfnamefont {B.}~\bibnamefont {Dennig}}, \bibinfo {author} {\bibfnamefont {S.}~\bibnamefont {Geisert}}, \bibinfo {author} {\bibfnamefont {S.}~\bibnamefont {Günzler}},
  \bibinfo {author} {\bibfnamefont {S.}~\bibnamefont {Ihssen}}, \bibinfo {author} {\bibfnamefont {P.}~\bibnamefont {Paluch}}, \bibinfo {author} {\bibfnamefont {T.}~\bibnamefont {Reisinger}}, \bibinfo {author} {\bibfnamefont {R.}~\bibnamefont {Hanna}}, \bibinfo {author} {\bibfnamefont {J.~H.}\ \bibnamefont {Bae}}, \bibinfo {author} {\bibfnamefont {P.}~\bibnamefont {Schüffelgen}}, \bibinfo {author} {\bibfnamefont {D.}~\bibnamefont {Grützmacher}}, \bibinfo {author} {\bibfnamefont {L.}~\bibnamefont {Buimaga-Iarinca}}, \bibinfo {author} {\bibfnamefont {C.}~\bibnamefont {Morari}}, \bibinfo {author} {\bibfnamefont {W.}~\bibnamefont {Wernsdorfer}}, \bibinfo {author} {\bibfnamefont {D.~P.}\ \bibnamefont {DiVincenzo}}, \bibinfo {author} {\bibfnamefont {K.}~\bibnamefont {Michielsen}}, \bibinfo {author} {\bibfnamefont {G.}~\bibnamefont {Catelani}},\ and\ \bibinfo {author} {\bibfnamefont {I.~M.}\ \bibnamefont {Pop}},\ }\href {https://doi.org/10.1038/s41567-024-02400-8} {\bibfield  {journal} {\bibinfo  {journal} {Nature
  Physics 2024}\ ,\ \bibinfo {pages} {1}} (\bibinfo {year} {2024})}\BibitemShut {NoStop}%
\bibitem [{\citenamefont {S\"ut\ifmmode~\mbox{\H{o}}\else \H{o}\fi{}}\ \emph {et~al.}(2022)\citenamefont {S\"ut\ifmmode~\mbox{\H{o}}\else \H{o}\fi{}}, \citenamefont {Prok}, \citenamefont {Makk}, \citenamefont {Kirti}, \citenamefont {Biasiol}, \citenamefont {Csonka},\ and\ \citenamefont {T\'ov\'ari}}]{Suto2022}%
  \BibitemOpen
  \bibfield  {author} {\bibinfo {author} {\bibfnamefont {M.}~\bibnamefont {S\"ut\ifmmode~\mbox{\H{o}}\else \H{o}\fi{}}}, \bibinfo {author} {\bibfnamefont {T.}~\bibnamefont {Prok}}, \bibinfo {author} {\bibfnamefont {P.}~\bibnamefont {Makk}}, \bibinfo {author} {\bibfnamefont {M.}~\bibnamefont {Kirti}}, \bibinfo {author} {\bibfnamefont {G.}~\bibnamefont {Biasiol}}, \bibinfo {author} {\bibfnamefont {S.}~\bibnamefont {Csonka}},\ and\ \bibinfo {author} {\bibfnamefont {E.}~\bibnamefont {T\'ov\'ari}},\ }\href {https://doi.org/10.1103/PhysRevB.106.235404} {\bibfield  {journal} {\bibinfo  {journal} {Phys. Rev. B}\ }\textbf {\bibinfo {volume} {106}},\ \bibinfo {pages} {235404} (\bibinfo {year} {2022})}\BibitemShut {NoStop}%
\bibitem [{\citenamefont {Benali}\ \emph {et~al.}(2022)\citenamefont {Benali}, \citenamefont {Rajak}, \citenamefont {Ciancio}, \citenamefont {Plaisier}, \citenamefont {Heun},\ and\ \citenamefont {Biasiol}}]{Benali2022}%
  \BibitemOpen
  \bibfield  {author} {\bibinfo {author} {\bibfnamefont {A.}~\bibnamefont {Benali}}, \bibinfo {author} {\bibfnamefont {P.}~\bibnamefont {Rajak}}, \bibinfo {author} {\bibfnamefont {R.}~\bibnamefont {Ciancio}}, \bibinfo {author} {\bibfnamefont {J.~R.}\ \bibnamefont {Plaisier}}, \bibinfo {author} {\bibfnamefont {S.}~\bibnamefont {Heun}},\ and\ \bibinfo {author} {\bibfnamefont {G.}~\bibnamefont {Biasiol}},\ }\href {https://doi.org/10.1016/J.JCRYSGRO.2022.126768} {\bibfield  {journal} {\bibinfo  {journal} {Journal of Crystal Growth}\ }\textbf {\bibinfo {volume} {593}},\ \bibinfo {pages} {126768} (\bibinfo {year} {2022})}\BibitemShut {NoStop}%
\bibitem [{\citenamefont {Khalil}\ \emph {et~al.}(2012)\citenamefont {Khalil}, \citenamefont {Stoutimore}, \citenamefont {Wellstood},\ and\ \citenamefont {Osborn}}]{khalil2012}%
  \BibitemOpen
  \bibfield  {author} {\bibinfo {author} {\bibfnamefont {M.~S.}\ \bibnamefont {Khalil}}, \bibinfo {author} {\bibfnamefont {M.~J.~A.}\ \bibnamefont {Stoutimore}}, \bibinfo {author} {\bibfnamefont {F.~C.}\ \bibnamefont {Wellstood}},\ and\ \bibinfo {author} {\bibfnamefont {K.~D.}\ \bibnamefont {Osborn}},\ }\href {https://doi.org/10.1063/1.3692073} {\bibfield  {journal} {\bibinfo  {journal} {Journal of Applied Physics}\ }\textbf {\bibinfo {volume} {111}},\  (\bibinfo {year} {2012})}\BibitemShut {NoStop}%
\bibitem [{\citenamefont {Probst}\ \emph {et~al.}(2015)\citenamefont {Probst}, \citenamefont {Song}, \citenamefont {Bushev}, \citenamefont {Ustinov},\ and\ \citenamefont {Weides}}]{Probst2015}%
  \BibitemOpen
  \bibfield  {author} {\bibinfo {author} {\bibfnamefont {S.}~\bibnamefont {Probst}}, \bibinfo {author} {\bibfnamefont {F.~B.}\ \bibnamefont {Song}}, \bibinfo {author} {\bibfnamefont {P.~A.}\ \bibnamefont {Bushev}}, \bibinfo {author} {\bibfnamefont {A.~V.}\ \bibnamefont {Ustinov}},\ and\ \bibinfo {author} {\bibfnamefont {M.}~\bibnamefont {Weides}},\ }\bibfield  {journal} {\bibinfo  {journal} {Review of Scientific Instruments}\ }\textbf {\bibinfo {volume} {86}},\ \href {https://doi.org/10.1063/1.4907935/360955} {10.1063/1.4907935/360955} (\bibinfo {year} {2015})\BibitemShut {NoStop}%
\bibitem [{\citenamefont {Wang}\ \emph {et~al.}(2021)\citenamefont {Wang}, \citenamefont {Singh}, \citenamefont {McRae}, \citenamefont {Bardin}, \citenamefont {Lin}, \citenamefont {Messaoudi}, \citenamefont {Castelli}, \citenamefont {Rosen}, \citenamefont {Holland}, \citenamefont {Pappas},\ and\ \citenamefont {Mutus}}]{Wang2021}%
  \BibitemOpen
  \bibfield  {author} {\bibinfo {author} {\bibfnamefont {H.}~\bibnamefont {Wang}}, \bibinfo {author} {\bibfnamefont {S.}~\bibnamefont {Singh}}, \bibinfo {author} {\bibfnamefont {C.~R.}\ \bibnamefont {McRae}}, \bibinfo {author} {\bibfnamefont {J.~C.}\ \bibnamefont {Bardin}}, \bibinfo {author} {\bibfnamefont {S.~X.}\ \bibnamefont {Lin}}, \bibinfo {author} {\bibfnamefont {N.}~\bibnamefont {Messaoudi}}, \bibinfo {author} {\bibfnamefont {A.~R.}\ \bibnamefont {Castelli}}, \bibinfo {author} {\bibfnamefont {Y.~J.}\ \bibnamefont {Rosen}}, \bibinfo {author} {\bibfnamefont {E.~T.}\ \bibnamefont {Holland}}, \bibinfo {author} {\bibfnamefont {D.~P.}\ \bibnamefont {Pappas}},\ and\ \bibinfo {author} {\bibfnamefont {J.~Y.}\ \bibnamefont {Mutus}},\ }\href {https://doi.org/10.1088/2058-9565/AC070E} {\bibfield  {journal} {\bibinfo  {journal} {Quantum Science and Technology}\ }\textbf {\bibinfo {volume} {6}},\ \bibinfo {pages} {035015} (\bibinfo {year} {2021})}\BibitemShut {NoStop}%
\bibitem [{\citenamefont {Strambini}\ \emph {et~al.}(2020)\citenamefont {Strambini}, \citenamefont {Iorio}, \citenamefont {Durante}, \citenamefont {Citro}, \citenamefont {Sanz-Fernández}, \citenamefont {Guarcello}, \citenamefont {Tokatly}, \citenamefont {Braggio}, \citenamefont {Rocci}, \citenamefont {Ligato}, \citenamefont {Zannier}, \citenamefont {Sorba}, \citenamefont {Bergeret},\ and\ \citenamefont {Giazotto}}]{Strambini2020}%
  \BibitemOpen
  \bibfield  {author} {\bibinfo {author} {\bibfnamefont {E.}~\bibnamefont {Strambini}}, \bibinfo {author} {\bibfnamefont {A.}~\bibnamefont {Iorio}}, \bibinfo {author} {\bibfnamefont {O.}~\bibnamefont {Durante}}, \bibinfo {author} {\bibfnamefont {R.}~\bibnamefont {Citro}}, \bibinfo {author} {\bibfnamefont {C.}~\bibnamefont {Sanz-Fernández}}, \bibinfo {author} {\bibfnamefont {C.}~\bibnamefont {Guarcello}}, \bibinfo {author} {\bibfnamefont {I.~V.}\ \bibnamefont {Tokatly}}, \bibinfo {author} {\bibfnamefont {A.}~\bibnamefont {Braggio}}, \bibinfo {author} {\bibfnamefont {M.}~\bibnamefont {Rocci}}, \bibinfo {author} {\bibfnamefont {N.}~\bibnamefont {Ligato}}, \bibinfo {author} {\bibfnamefont {V.}~\bibnamefont {Zannier}}, \bibinfo {author} {\bibfnamefont {L.}~\bibnamefont {Sorba}}, \bibinfo {author} {\bibfnamefont {F.~S.}\ \bibnamefont {Bergeret}},\ and\ \bibinfo {author} {\bibfnamefont {F.}~\bibnamefont {Giazotto}},\ }\href {https://doi.org/10.1038/s41565-020-0712-7} {\bibfield  {journal} {\bibinfo  {journal}
  {Nature Nanotechnology 2020 15:8}\ }\textbf {\bibinfo {volume} {15}},\ \bibinfo {pages} {656} (\bibinfo {year} {2020})}\BibitemShut {NoStop}%
\bibitem [{\citenamefont {Virtanen}\ \emph {et~al.}(2011)\citenamefont {Virtanen}, \citenamefont {Bergeret}, \citenamefont {Cuevas},\ and\ \citenamefont {Heikkil\"a}}]{Virtanen2011}%
  \BibitemOpen
  \bibfield  {author} {\bibinfo {author} {\bibfnamefont {P.}~\bibnamefont {Virtanen}}, \bibinfo {author} {\bibfnamefont {F.~S.}\ \bibnamefont {Bergeret}}, \bibinfo {author} {\bibfnamefont {J.~C.}\ \bibnamefont {Cuevas}},\ and\ \bibinfo {author} {\bibfnamefont {T.~T.}\ \bibnamefont {Heikkil\"a}},\ }\href {https://doi.org/10.1103/PhysRevB.83.144514} {\bibfield  {journal} {\bibinfo  {journal} {Phys. Rev. B}\ }\textbf {\bibinfo {volume} {83}},\ \bibinfo {pages} {144514} (\bibinfo {year} {2011})}\BibitemShut {NoStop}%
\bibitem [{\citenamefont {McRae}\ \emph {et~al.}(2020)\citenamefont {McRae}, \citenamefont {Wang}, \citenamefont {Gao}, \citenamefont {Vissers}, \citenamefont {Brecht}, \citenamefont {Dunsworth}, \citenamefont {Pappas},\ and\ \citenamefont {Mutus}}]{Corey2020}%
  \BibitemOpen
  \bibfield  {author} {\bibinfo {author} {\bibfnamefont {C.~R.~H.}\ \bibnamefont {McRae}}, \bibinfo {author} {\bibfnamefont {H.}~\bibnamefont {Wang}}, \bibinfo {author} {\bibfnamefont {J.}~\bibnamefont {Gao}}, \bibinfo {author} {\bibfnamefont {M.}~\bibnamefont {Vissers}}, \bibinfo {author} {\bibfnamefont {T.}~\bibnamefont {Brecht}}, \bibinfo {author} {\bibfnamefont {A.}~\bibnamefont {Dunsworth}}, \bibinfo {author} {\bibfnamefont {D.}~\bibnamefont {Pappas}},\ and\ \bibinfo {author} {\bibfnamefont {J.}~\bibnamefont {Mutus}},\ }\href {https://aip.scitation.org/doi/10.1063/5.0017378} {\bibfield  {journal} {\bibinfo  {journal} {Review of Scientific Instruments (invited)}\ } (\bibinfo {year} {2020})}\BibitemShut {NoStop}%
\bibitem [{\citenamefont {Healey}\ \emph {et~al.}(2008)\citenamefont {Healey}, \citenamefont {Lindström}, \citenamefont {Colclough}, \citenamefont {Muirhead},\ and\ \citenamefont {Tzalenchuk}}]{Healey2008}%
  \BibitemOpen
  \bibfield  {author} {\bibinfo {author} {\bibfnamefont {J.~E.}\ \bibnamefont {Healey}}, \bibinfo {author} {\bibfnamefont {T.}~\bibnamefont {Lindström}}, \bibinfo {author} {\bibfnamefont {M.~S.}\ \bibnamefont {Colclough}}, \bibinfo {author} {\bibfnamefont {C.~M.}\ \bibnamefont {Muirhead}},\ and\ \bibinfo {author} {\bibfnamefont {A.~Y.}\ \bibnamefont {Tzalenchuk}},\ }\href {https://doi.org/10.1063/1.2959824} {\bibfield  {journal} {\bibinfo  {journal} {Applied Physics Letters}\ }\textbf {\bibinfo {volume} {93}},\ \bibinfo {pages} {043513} (\bibinfo {year} {2008})}\BibitemShut {NoStop}%
\bibitem [{\citenamefont {Samkharadze}\ \emph {et~al.}(2016)\citenamefont {Samkharadze}, \citenamefont {Bruno}, \citenamefont {Scarlino}, \citenamefont {Zheng}, \citenamefont {DiVincenzo}, \citenamefont {DiCarlo},\ and\ \citenamefont {Vandersypen}}]{Samkharadze2016}%
  \BibitemOpen
  \bibfield  {author} {\bibinfo {author} {\bibfnamefont {N.}~\bibnamefont {Samkharadze}}, \bibinfo {author} {\bibfnamefont {A.}~\bibnamefont {Bruno}}, \bibinfo {author} {\bibfnamefont {P.}~\bibnamefont {Scarlino}}, \bibinfo {author} {\bibfnamefont {G.}~\bibnamefont {Zheng}}, \bibinfo {author} {\bibfnamefont {D.~P.}\ \bibnamefont {DiVincenzo}}, \bibinfo {author} {\bibfnamefont {L.}~\bibnamefont {DiCarlo}},\ and\ \bibinfo {author} {\bibfnamefont {L.~M.~K.}\ \bibnamefont {Vandersypen}},\ }\href {https://doi.org/10.1103/PhysRevApplied.5.044004} {\bibfield  {journal} {\bibinfo  {journal} {Phys. Rev. Appl.}\ }\textbf {\bibinfo {volume} {5}},\ \bibinfo {pages} {044004} (\bibinfo {year} {2016})}\BibitemShut {NoStop}%
\bibitem [{\citenamefont {L\'opez-N\'u\~nez}\ \emph {et~al.}(2023)\citenamefont {L\'opez-N\'u\~nez}, \citenamefont {Montserrat}, \citenamefont {Rius}, \citenamefont {Bertoldo}, \citenamefont {Torras-Coloma}, \citenamefont {Mart\'\i{}nez},\ and\ \citenamefont {Forn-D\'\i{}az}}]{Lopez-Nunez2023}%
  \BibitemOpen
  \bibfield  {author} {\bibinfo {author} {\bibfnamefont {D.}~\bibnamefont {L\'opez-N\'u\~nez}}, \bibinfo {author} {\bibfnamefont {Q.~P.}\ \bibnamefont {Montserrat}}, \bibinfo {author} {\bibfnamefont {G.}~\bibnamefont {Rius}}, \bibinfo {author} {\bibfnamefont {E.}~\bibnamefont {Bertoldo}}, \bibinfo {author} {\bibfnamefont {A.}~\bibnamefont {Torras-Coloma}}, \bibinfo {author} {\bibfnamefont {M.}~\bibnamefont {Mart\'\i{}nez}},\ and\ \bibinfo {author} {\bibfnamefont {P.}~\bibnamefont {Forn-D\'\i{}az}},\ }\href@noop {} {\bibfield  {journal} {\bibinfo  {journal} {ArXiv}\ } (\bibinfo {year} {2023})},\ \Eprint {https://arxiv.org/abs/2311.14119} {arXiv:2311.14119 [cond-mat.supr-con]} \BibitemShut {NoStop}%
\bibitem [{\citenamefont {Bothner}\ \emph {et~al.}(2012)\citenamefont {Bothner}, \citenamefont {Gaber}, \citenamefont {Kemmler}, \citenamefont {Koelle}, \citenamefont {Kleiner}, \citenamefont {W\"unsch},\ and\ \citenamefont {Siegel}}]{Bothner2012}%
  \BibitemOpen
  \bibfield  {author} {\bibinfo {author} {\bibfnamefont {D.}~\bibnamefont {Bothner}}, \bibinfo {author} {\bibfnamefont {T.}~\bibnamefont {Gaber}}, \bibinfo {author} {\bibfnamefont {M.}~\bibnamefont {Kemmler}}, \bibinfo {author} {\bibfnamefont {D.}~\bibnamefont {Koelle}}, \bibinfo {author} {\bibfnamefont {R.}~\bibnamefont {Kleiner}}, \bibinfo {author} {\bibfnamefont {S.}~\bibnamefont {W\"unsch}},\ and\ \bibinfo {author} {\bibfnamefont {M.}~\bibnamefont {Siegel}},\ }\href {https://doi.org/10.1103/PhysRevB.86.014517} {\bibfield  {journal} {\bibinfo  {journal} {Phys. Rev. B}\ }\textbf {\bibinfo {volume} {86}},\ \bibinfo {pages} {014517} (\bibinfo {year} {2012})}\BibitemShut {NoStop}%
\bibitem [{\citenamefont {R.}\ and\ \citenamefont {B.}(1969)}]{Merservey1969}%
  \BibitemOpen
  \bibfield  {author} {\bibinfo {author} {\bibfnamefont {M.}~\bibnamefont {R.}}\ and\ \bibinfo {author} {\bibfnamefont {S.~B.}\ \bibnamefont {B.}},\ }\href@noop {} {\emph {\bibinfo {title} {Superconductivity}}},\ edited by\ \bibinfo {editor} {\bibfnamefont {R.~D.}\ \bibnamefont {Parks}}\ (\bibinfo  {publisher} {Marcel Dekker, Inc.},\ \bibinfo {year} {1969})\ p.\ \bibinfo {pages} {126}\BibitemShut {NoStop}%
\bibitem [{\citenamefont {Pearl}(1964)}]{Pearl1964}%
  \BibitemOpen
  \bibfield  {author} {\bibinfo {author} {\bibfnamefont {J.}~\bibnamefont {Pearl}},\ }\href {https://doi.org/10.1063/1.1754056} {\bibfield  {journal} {\bibinfo  {journal} {Applied Physics Letters}\ }\textbf {\bibinfo {volume} {5}},\ \bibinfo {pages} {65} (\bibinfo {year} {1964})}\BibitemShut {NoStop}%
\bibitem [{\citenamefont {Zeldov}\ \emph {et~al.}(1994)\citenamefont {Zeldov}, \citenamefont {Clem}, \citenamefont {McElfresh},\ and\ \citenamefont {Darwin}}]{Zeldov1994}%
  \BibitemOpen
  \bibfield  {author} {\bibinfo {author} {\bibfnamefont {E.}~\bibnamefont {Zeldov}}, \bibinfo {author} {\bibfnamefont {J.~R.}\ \bibnamefont {Clem}}, \bibinfo {author} {\bibfnamefont {M.}~\bibnamefont {McElfresh}},\ and\ \bibinfo {author} {\bibfnamefont {M.}~\bibnamefont {Darwin}},\ }\href {https://doi.org/10.1103/PhysRevB.49.9802} {\bibfield  {journal} {\bibinfo  {journal} {Phys. Rev. B}\ }\textbf {\bibinfo {volume} {49}},\ \bibinfo {pages} {9802} (\bibinfo {year} {1994})}\BibitemShut {NoStop}%
\bibitem [{\citenamefont {Mayer}\ \emph {et~al.}(2020)\citenamefont {Mayer}, \citenamefont {Dartiailh}, \citenamefont {Yuan}, \citenamefont {Wickramasinghe}, \citenamefont {Rossi},\ and\ \citenamefont {Shabani}}]{Mayer2020}%
  \BibitemOpen
  \bibfield  {author} {\bibinfo {author} {\bibfnamefont {W.}~\bibnamefont {Mayer}}, \bibinfo {author} {\bibfnamefont {M.~C.}\ \bibnamefont {Dartiailh}}, \bibinfo {author} {\bibfnamefont {J.}~\bibnamefont {Yuan}}, \bibinfo {author} {\bibfnamefont {K.~S.}\ \bibnamefont {Wickramasinghe}}, \bibinfo {author} {\bibfnamefont {E.}~\bibnamefont {Rossi}},\ and\ \bibinfo {author} {\bibfnamefont {J.}~\bibnamefont {Shabani}},\ }\href {https://doi.org/10.1038/s41467-019-14094-1} {\bibfield  {journal} {\bibinfo  {journal} {Nature Communications 2020 11:1}\ }\textbf {\bibinfo {volume} {11}},\ \bibinfo {pages} {1} (\bibinfo {year} {2020})}\BibitemShut {NoStop}%
\bibitem [{\citenamefont {Lotfizadeh}\ \emph {et~al.}(2024)\citenamefont {Lotfizadeh}, \citenamefont {Schiela}, \citenamefont {Pekerten}, \citenamefont {Yu}, \citenamefont {Elfeky}, \citenamefont {Strickland}, \citenamefont {Matos-Abiague},\ and\ \citenamefont {Shabani}}]{Lotfizadeh2024}%
  \BibitemOpen
  \bibfield  {author} {\bibinfo {author} {\bibfnamefont {N.}~\bibnamefont {Lotfizadeh}}, \bibinfo {author} {\bibfnamefont {W.~F.}\ \bibnamefont {Schiela}}, \bibinfo {author} {\bibfnamefont {B.}~\bibnamefont {Pekerten}}, \bibinfo {author} {\bibfnamefont {P.}~\bibnamefont {Yu}}, \bibinfo {author} {\bibfnamefont {B.~H.}\ \bibnamefont {Elfeky}}, \bibinfo {author} {\bibfnamefont {W.~M.}\ \bibnamefont {Strickland}}, \bibinfo {author} {\bibfnamefont {A.}~\bibnamefont {Matos-Abiague}},\ and\ \bibinfo {author} {\bibfnamefont {J.}~\bibnamefont {Shabani}},\ }\href {https://doi.org/10.1038/s42005-024-01618-5} {\bibfield  {journal} {\bibinfo  {journal} {Communications Physics 2024 7:1}\ }\textbf {\bibinfo {volume} {7}},\ \bibinfo {pages} {1} (\bibinfo {year} {2024})}\BibitemShut {NoStop}%
\bibitem [{\citenamefont {Baumgartner}\ \emph {et~al.}(2021)\citenamefont {Baumgartner}, \citenamefont {Fuchs}, \citenamefont {Costa}, \citenamefont {Reinhardt}, \citenamefont {Gronin}, \citenamefont {Gardner}, \citenamefont {Lindemann}, \citenamefont {Manfra}, \citenamefont {Junior}, \citenamefont {Kochan}, \citenamefont {Fabian}, \citenamefont {Paradiso},\ and\ \citenamefont {Strunk}}]{Baumgartner2021}%
  \BibitemOpen
  \bibfield  {author} {\bibinfo {author} {\bibfnamefont {C.}~\bibnamefont {Baumgartner}}, \bibinfo {author} {\bibfnamefont {L.}~\bibnamefont {Fuchs}}, \bibinfo {author} {\bibfnamefont {A.}~\bibnamefont {Costa}}, \bibinfo {author} {\bibfnamefont {S.}~\bibnamefont {Reinhardt}}, \bibinfo {author} {\bibfnamefont {S.}~\bibnamefont {Gronin}}, \bibinfo {author} {\bibfnamefont {G.~C.}\ \bibnamefont {Gardner}}, \bibinfo {author} {\bibfnamefont {T.}~\bibnamefont {Lindemann}}, \bibinfo {author} {\bibfnamefont {M.~J.}\ \bibnamefont {Manfra}}, \bibinfo {author} {\bibfnamefont {P.~E.~F.}\ \bibnamefont {Junior}}, \bibinfo {author} {\bibfnamefont {D.}~\bibnamefont {Kochan}}, \bibinfo {author} {\bibfnamefont {J.}~\bibnamefont {Fabian}}, \bibinfo {author} {\bibfnamefont {N.}~\bibnamefont {Paradiso}},\ and\ \bibinfo {author} {\bibfnamefont {C.}~\bibnamefont {Strunk}},\ }\href {https://doi.org/10.1038/s41565-021-01009-9} {\bibfield  {journal} {\bibinfo  {journal} {Nature Nanotechnology 2021 17:1}\ }\textbf {\bibinfo {volume}
  {17}},\ \bibinfo {pages} {39} (\bibinfo {year} {2021})}\BibitemShut {NoStop}%
\bibitem [{\citenamefont {Reinhardt}\ \emph {et~al.}(2024)\citenamefont {Reinhardt}, \citenamefont {Ascherl}, \citenamefont {Costa}, \citenamefont {Berger}, \citenamefont {Gronin}, \citenamefont {Gardner}, \citenamefont {Lindemann}, \citenamefont {Manfra}, \citenamefont {Fabian}, \citenamefont {Kochan}, \citenamefont {Strunk},\ and\ \citenamefont {Paradiso}}]{Reinhardt2024}%
  \BibitemOpen
  \bibfield  {author} {\bibinfo {author} {\bibfnamefont {S.}~\bibnamefont {Reinhardt}}, \bibinfo {author} {\bibfnamefont {T.}~\bibnamefont {Ascherl}}, \bibinfo {author} {\bibfnamefont {A.}~\bibnamefont {Costa}}, \bibinfo {author} {\bibfnamefont {J.}~\bibnamefont {Berger}}, \bibinfo {author} {\bibfnamefont {S.}~\bibnamefont {Gronin}}, \bibinfo {author} {\bibfnamefont {G.~C.}\ \bibnamefont {Gardner}}, \bibinfo {author} {\bibfnamefont {T.}~\bibnamefont {Lindemann}}, \bibinfo {author} {\bibfnamefont {M.~J.}\ \bibnamefont {Manfra}}, \bibinfo {author} {\bibfnamefont {J.}~\bibnamefont {Fabian}}, \bibinfo {author} {\bibfnamefont {D.}~\bibnamefont {Kochan}}, \bibinfo {author} {\bibfnamefont {C.}~\bibnamefont {Strunk}},\ and\ \bibinfo {author} {\bibfnamefont {N.}~\bibnamefont {Paradiso}},\ }\href {https://doi.org/10.1038/s41467-024-48741-z} {\bibfield  {journal} {\bibinfo  {journal} {Nature Communications 2024 15:1}\ }\textbf {\bibinfo {volume} {15}},\ \bibinfo {pages} {1} (\bibinfo {year} {2024})}\BibitemShut
  {NoStop}%
\bibitem [{\citenamefont {Suominen}\ \emph {et~al.}(2017)\citenamefont {Suominen}, \citenamefont {Danon}, \citenamefont {Kjaergaard}, \citenamefont {Flensberg}, \citenamefont {Shabani}, \citenamefont {Palmstr\o{}m}, \citenamefont {Nichele},\ and\ \citenamefont {Marcus}}]{Suominen2017}%
  \BibitemOpen
  \bibfield  {author} {\bibinfo {author} {\bibfnamefont {H.~J.}\ \bibnamefont {Suominen}}, \bibinfo {author} {\bibfnamefont {J.}~\bibnamefont {Danon}}, \bibinfo {author} {\bibfnamefont {M.}~\bibnamefont {Kjaergaard}}, \bibinfo {author} {\bibfnamefont {K.}~\bibnamefont {Flensberg}}, \bibinfo {author} {\bibfnamefont {J.}~\bibnamefont {Shabani}}, \bibinfo {author} {\bibfnamefont {C.~J.}\ \bibnamefont {Palmstr\o{}m}}, \bibinfo {author} {\bibfnamefont {F.}~\bibnamefont {Nichele}},\ and\ \bibinfo {author} {\bibfnamefont {C.~M.}\ \bibnamefont {Marcus}},\ }\href {https://doi.org/10.1103/PhysRevB.95.035307} {\bibfield  {journal} {\bibinfo  {journal} {Phys. Rev. B}\ }\textbf {\bibinfo {volume} {95}},\ \bibinfo {pages} {035307} (\bibinfo {year} {2017})}\BibitemShut {NoStop}%
\bibitem [{\citenamefont {Assouline}\ \emph {et~al.}(2019)\citenamefont {Assouline}, \citenamefont {Feuillet-Palma}, \citenamefont {Bergeal}, \citenamefont {Zhang}, \citenamefont {Mottaghizadeh}, \citenamefont {Zimmers}, \citenamefont {Lhuillier}, \citenamefont {Eddrie}, \citenamefont {Atkinson}, \citenamefont {Aprili},\ and\ \citenamefont {Aubin}}]{Assouline2019}%
  \BibitemOpen
  \bibfield  {author} {\bibinfo {author} {\bibfnamefont {A.}~\bibnamefont {Assouline}}, \bibinfo {author} {\bibfnamefont {C.}~\bibnamefont {Feuillet-Palma}}, \bibinfo {author} {\bibfnamefont {N.}~\bibnamefont {Bergeal}}, \bibinfo {author} {\bibfnamefont {T.}~\bibnamefont {Zhang}}, \bibinfo {author} {\bibfnamefont {A.}~\bibnamefont {Mottaghizadeh}}, \bibinfo {author} {\bibfnamefont {A.}~\bibnamefont {Zimmers}}, \bibinfo {author} {\bibfnamefont {E.}~\bibnamefont {Lhuillier}}, \bibinfo {author} {\bibfnamefont {M.}~\bibnamefont {Eddrie}}, \bibinfo {author} {\bibfnamefont {P.}~\bibnamefont {Atkinson}}, \bibinfo {author} {\bibfnamefont {M.}~\bibnamefont {Aprili}},\ and\ \bibinfo {author} {\bibfnamefont {H.}~\bibnamefont {Aubin}},\ }\href {https://doi.org/10.1038/s41467-018-08022-y} {\bibfield  {journal} {\bibinfo  {journal} {Nature Communications 2019 10:1}\ }\textbf {\bibinfo {volume} {10}},\ \bibinfo {pages} {1} (\bibinfo {year} {2019})}\BibitemShut {NoStop}%
\bibitem [{\citenamefont {Banerjee}\ \emph {et~al.}(2023)\citenamefont {Banerjee}, \citenamefont {Lesser}, \citenamefont {Rahman}, \citenamefont {Thomas}, \citenamefont {Wang}, \citenamefont {Manfra}, \citenamefont {Berg}, \citenamefont {Oreg}, \citenamefont {Stern},\ and\ \citenamefont {Marcus}}]{Banerjee2023}%
  \BibitemOpen
  \bibfield  {author} {\bibinfo {author} {\bibfnamefont {A.}~\bibnamefont {Banerjee}}, \bibinfo {author} {\bibfnamefont {O.}~\bibnamefont {Lesser}}, \bibinfo {author} {\bibfnamefont {M.~A.}\ \bibnamefont {Rahman}}, \bibinfo {author} {\bibfnamefont {C.}~\bibnamefont {Thomas}}, \bibinfo {author} {\bibfnamefont {T.}~\bibnamefont {Wang}}, \bibinfo {author} {\bibfnamefont {M.~J.}\ \bibnamefont {Manfra}}, \bibinfo {author} {\bibfnamefont {E.}~\bibnamefont {Berg}}, \bibinfo {author} {\bibfnamefont {Y.}~\bibnamefont {Oreg}}, \bibinfo {author} {\bibfnamefont {A.}~\bibnamefont {Stern}},\ and\ \bibinfo {author} {\bibfnamefont {C.~M.}\ \bibnamefont {Marcus}},\ }\href {https://doi.org/10.1103/PhysRevLett.130.096202} {\bibfield  {journal} {\bibinfo  {journal} {Phys. Rev. Lett.}\ }\textbf {\bibinfo {volume} {130}},\ \bibinfo {pages} {096202} (\bibinfo {year} {2023})}\BibitemShut {NoStop}%
\end{thebibliography}%


\begin{thebibliography}{3}%
\makeatletter
\providecommand \@ifxundefined [1]{%
 \@ifx{#1\undefined}
}%
\providecommand \@ifnum [1]{%
 \ifnum #1\expandafter \@firstoftwo
 \else \expandafter \@secondoftwo
 \fi
}%
\providecommand \@ifx [1]{%
 \ifx #1\expandafter \@firstoftwo
 \else \expandafter \@secondoftwo
 \fi
}%
\providecommand \natexlab [1]{#1}%
\providecommand \enquote  [1]{``#1''}%
\providecommand \bibnamefont  [1]{#1}%
\providecommand \bibfnamefont [1]{#1}%
\providecommand \citenamefont [1]{#1}%
\providecommand \href@noop [0]{\@secondoftwo}%
\providecommand \href [0]{\begingroup \@sanitize@url \@href}%
\providecommand \@href[1]{\@@startlink{#1}\@@href}%
\providecommand \@@href[1]{\endgroup#1\@@endlink}%
\providecommand \@sanitize@url [0]{\catcode `\\12\catcode `\$12\catcode `\&12\catcode `\#12\catcode `\^12\catcode `\_12\catcode `\%12\relax}%
\providecommand \@@startlink[1]{}%
\providecommand \@@endlink[0]{}%
\providecommand \url  [0]{\begingroup\@sanitize@url \@url }%
\providecommand \@url [1]{\endgroup\@href {#1}{\urlprefix }}%
\providecommand \urlprefix  [0]{URL }%
\providecommand \Eprint [0]{\href }%
\providecommand \doibase [0]{https://doi.org/}%
\providecommand \selectlanguage [0]{\@gobble}%
\providecommand \bibinfo  [0]{\@secondoftwo}%
\providecommand \bibfield  [0]{\@secondoftwo}%
\providecommand \translation [1]{[#1]}%
\providecommand \BibitemOpen [0]{}%
\providecommand \bibitemStop [0]{}%
\providecommand \bibitemNoStop [0]{.\EOS\space}%
\providecommand \EOS [0]{\spacefactor3000\relax}%
\providecommand \BibitemShut  [1]{\csname bibitem#1\endcsname}%
\let\auto@bib@innerbib\@empty
\bibitem [{\citenamefont {S\"ut\ifmmode~\mbox{\H{o}}\else \H{o}\fi{}}\ \emph {et~al.}(2022)\citenamefont {S\"ut\ifmmode~\mbox{\H{o}}\else \H{o}\fi{}}, \citenamefont {Prok}, \citenamefont {Makk}, \citenamefont {Kirti}, \citenamefont {Biasiol}, \citenamefont {Csonka},\ and\ \citenamefont {T\'ov\'ari}}]{Suto2022}%
  \BibitemOpen
  \bibfield  {author} {\bibinfo {author} {\bibfnamefont {M.}~\bibnamefont {S\"ut\ifmmode~\mbox{\H{o}}\else \H{o}\fi{}}}, \bibinfo {author} {\bibfnamefont {T.}~\bibnamefont {Prok}}, \bibinfo {author} {\bibfnamefont {P.}~\bibnamefont {Makk}}, \bibinfo {author} {\bibfnamefont {M.}~\bibnamefont {Kirti}}, \bibinfo {author} {\bibfnamefont {G.}~\bibnamefont {Biasiol}}, \bibinfo {author} {\bibfnamefont {S.}~\bibnamefont {Csonka}},\ and\ \bibinfo {author} {\bibfnamefont {E.}~\bibnamefont {T\'ov\'ari}},\ }\href {https://doi.org/10.1103/PhysRevB.106.235404} {\bibfield  {journal} {\bibinfo  {journal} {Phys. Rev. B}\ }\textbf {\bibinfo {volume} {106}},\ \bibinfo {pages} {235404} (\bibinfo {year} {2022})}\BibitemShut {NoStop}%
\bibitem [{\citenamefont {Benali}\ \emph {et~al.}(2022)\citenamefont {Benali}, \citenamefont {Rajak}, \citenamefont {Ciancio}, \citenamefont {Plaisier}, \citenamefont {Heun},\ and\ \citenamefont {Biasiol}}]{Benali2022}%
  \BibitemOpen
  \bibfield  {author} {\bibinfo {author} {\bibfnamefont {A.}~\bibnamefont {Benali}}, \bibinfo {author} {\bibfnamefont {P.}~\bibnamefont {Rajak}}, \bibinfo {author} {\bibfnamefont {R.}~\bibnamefont {Ciancio}}, \bibinfo {author} {\bibfnamefont {J.~R.}\ \bibnamefont {Plaisier}}, \bibinfo {author} {\bibfnamefont {S.}~\bibnamefont {Heun}},\ and\ \bibinfo {author} {\bibfnamefont {G.}~\bibnamefont {Biasiol}},\ }\href {https://doi.org/10.1016/J.JCRYSGRO.2022.126768} {\bibfield  {journal} {\bibinfo  {journal} {Journal of Crystal Growth}\ }\textbf {\bibinfo {volume} {593}},\ \bibinfo {pages} {126768} (\bibinfo {year} {2022})}\BibitemShut {NoStop}%
\bibitem [{\citenamefont {Suominen}\ \emph {et~al.}(2017)\citenamefont {Suominen}, \citenamefont {Danon}, \citenamefont {Kjaergaard}, \citenamefont {Flensberg}, \citenamefont {Shabani}, \citenamefont {Palmstr\o{}m}, \citenamefont {Nichele},\ and\ \citenamefont {Marcus}}]{Suominen2017}%
  \BibitemOpen
  \bibfield  {author} {\bibinfo {author} {\bibfnamefont {H.~J.}\ \bibnamefont {Suominen}}, \bibinfo {author} {\bibfnamefont {J.}~\bibnamefont {Danon}}, \bibinfo {author} {\bibfnamefont {M.}~\bibnamefont {Kjaergaard}}, \bibinfo {author} {\bibfnamefont {K.}~\bibnamefont {Flensberg}}, \bibinfo {author} {\bibfnamefont {J.}~\bibnamefont {Shabani}}, \bibinfo {author} {\bibfnamefont {C.~J.}\ \bibnamefont {Palmstr\o{}m}}, \bibinfo {author} {\bibfnamefont {F.}~\bibnamefont {Nichele}},\ and\ \bibinfo {author} {\bibfnamefont {C.~M.}\ \bibnamefont {Marcus}},\ }\href {https://doi.org/10.1103/PhysRevB.95.035307} {\bibfield  {journal} {\bibinfo  {journal} {Phys. Rev. B}\ }\textbf {\bibinfo {volume} {95}},\ \bibinfo {pages} {035307} (\bibinfo {year} {2017})}\BibitemShut {NoStop}%
\end{thebibliography}%
\end{document}


\widetext

\setcounter{equation}{0}
\setcounter{figure}{0}
\setcounter{table}{0}
\setcounter{page}{1}
\setcounter{section}{0}

\renewcommand{\thefigure}{S\arabic{figure}}
\renewcommand{\theequation}{S\arabic{equation}}
\renewcommand{\thesection}{S\Roman{section}}
\renewcommand{\bibnumfmt}[1]{[S#1]}
\renewcommand{\citenumfont}[1]{S#1}

\textbf{\centering\large Supplementary Material:\\ Determination of the current-phase relation of an InAs 2DEG Josephson junction with a microwave resonator}

\section{\label{sec:sample_fab}Sample fabrication}

The detailed layer composition of the GaAs-based semiconducting heterostructure wafer is published in Refs.\,\cite{Suto2022, Benali2022}, except here we used a 50-nm-thick epitaxial Al layer, while Ref.\,\cite{Suto2022} had a 10-nm-thick Al layer. We take advantage of the crystallographic features appearing in the optical microscope image under dark field illumination to identify the crystallographic directions\,\cite{Benali2022}. After the MBE growth, we fabricated the samples in the following steps.   

\medskip

\textbf{Forming the resonator and the SQUID loop, etching the mesa}
\begin{itemize}
    \item O$_2$ plasma cleaning using reactive ion etching
    \item Thin layer of adhesion promoter AR 300-80 NEW applied by vaporization
    \item Two layer of MAA resist, 4000 RPM 40$\,$s, baked after each at 185\textdegree C for 90$\,$s, for a combined thickness of 300$\,$nm 
    \item Electron beam exposure in a Raith150 system, 20$\,$kV, 80$\,$\textmu C/cm$^2$ dosage to create the resonators and the superconducting loops, without the Josephson junction. 
    \item Development in MIBK:IPA 1:3 for 60$\,$s, rinsed in IPA for 30$\,$s. 
    \item Al etching using MF-21, at 30\textdegree C for 80$\,$s while gently stirring with a magnetic stirrer. 
    \item Mesa etching using H$_2$O:C$_6$H$_8$O$_7$:H$_3$PO$_4$:H$_2$O$_2$ with a weight ratio of 220:55:3:3, at 30\textdegree C, gently stirred. The etching speed is around 200$\,$nm/min, the usual etch time is 2.5$\,$min to exceed 400$\,$nm. A shallower mesa etch causes leakage after high-temperature ALD. 
\end{itemize}

\textbf{Etching the Josephson junction}
 \begin{itemize}
    \item We continue the processing without removing the resist from the previous step
    \item Electron beam exposure and Al etching as before, to create the Josephson junction and remove the outreaching Al ledges created during the mesa etching. 
    \item Removal of the resist with aceton, rinsed in IPA
    \item O$_2$ plasma cleaning using reactive ion etching
\end{itemize}

\textbf{Creating the gate electrode}
\begin{itemize}
    \item Atomic layer deposition (ALD) of 50$\,$nm Al$_2$O$_3$ at 225\textdegree C using a PicoSun system. 
    \item 600K PMMA 4000 RPM 40$\,$s with 3\,min 170\textdegree C baking for a 300$\,$nm layer. 
    \item EBL 20$\,$kV, 240$\,$\textmu C/cm$^2$ for the gate electrodes 
    \item Metal deposition of Ti/Au of 20/180$\,$nm at 1$\,$\AA/s and 3$\,$\AA/s respectively, using a 45\textdegree  ~sample holder for device~A, and 10/90$\,$nm without the tilted sample holder for device B 
    \item Lift-off process in hot acetone, sonication if needed. Rinsed in IPA.
\end{itemize}

Mesa etching exhibits a strong asymmetry, which can be taken advantage of to make the tilted evaporation unnecessary, if the geometry is aligned with the crystalline directions of the wafer. The [$\overline{1}10$] direction gives a gentle slope, allowing the use of thinner metal gates. 

\medskip

\textbf{Microwave design}

\medskip

The coplanar transmission line used in the feedline and the resonator structure both have a characteristic impedance of $\sim50$\,$\Omega$, defined by the center conductor line of width $S=15$\,\textmu m and gap size $W=10.4$\,\textmu m. We found that the kinetic inductance of the Al layer is negligible at the base temperature, in $B=0$. The capacitance and inductance of the equivalent lumped resonator are $ C_p  ={\mathcal{C}_r l}/{2}$ and $ L_p  ={2}/{(l \omega_0^2 \mathcal{C}_r)}$, where $\omega_0$ is the resonant angular frequency, $l$ is the length and $\mathcal{C}_r$ is capacitance per unit length of the CPW transmission line. We modeled the circuit area around the rf SQUID loop in Ansys HFSS to get the resonator-SQUID coupling, and obtained $M=20.8$\,pH for the mutual inductance, and $L_{\mathrm{loop}}=154.5$\,pH for the self-inductance of the SQUID loop.

\section{\label{sec:setup}Measurement setup}
\begin{figure}[h]
\includegraphics{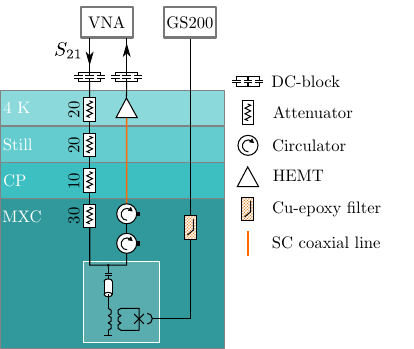}
\caption{\label{setup_fig} Measurement setup. The blue boxes indicate different stages of the dilution refrigerator. Attenuations on the input line are shown in dB. The chip is enclosed in a white rectangle.}
\end{figure}

In Fig.\,\ref{setup_fig} we show the measurement setup for the low-temperature experiments. The sample was cooled down in a Leiden Cryogenics CF-CS110-450 dilution refrigerator. We measured the microwave scattering parameter $S_{12}$ with a Rohde\&Schwarz ZNB20 vector network analyzer (VNA). We supplied the gate voltage with a Yokogawa GS200. On the high-frequency output line, we employed a Low Noise Factory dual junction circulator type LNF-CICIC4\_12A and a HEMT amplifier LNF-LNC4\_16B. We applied the magnetic field using the 3D vector magnet of the cryostat, we used a Yokogawa GS200 in current source mode to create the out-of-plane component, and an American Magnetics Inc. 4Q06125PS-430 integrated magnet power supply for the in-plane component.

\section{\label{sec:eval}Data evaluation}

Before fitting the resonances in $S_{12}$, we corrected for the standing-wave background originating from the imperfections of the microwave environment and accounted for the attenuation of the measurement chain in the following way. We applied a magnetic field $B_z$ of 2--3\,mT to suppress the superconductivity of the Al resonator to get the background transmission. The raw $S_{12}$ curves were normalized by this background curve, and we further corrected the magnitude by $\sim0.4\%$ to ensure $|S_{12}|=1$ far from resonance.

\begin{figure}[htb]
\includegraphics{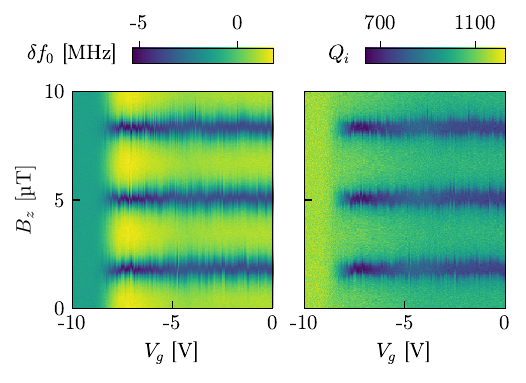}
\caption{\label{supp_f0map_Qimap} Gate and magnetic field dependence of $\delta f_0$ and $Q_i$.}
\end{figure}

For the determination of the current-phase relation, the resonance curves in the microwave transmission have been fitted with Equation 1 in two passes. In the first pass all fit parameters ($f_0$, $Q_i$, $Q_c$, $\vartheta$) were free fit parameters. With the assumption that the coupling quality factor $Q_c$ does not depend on the gate voltage $V_g$ nor the magnetic field $B_z$ (on the order of a few \textmu T), we calculated its mean value $Q_{c,\mathrm{avg}}$ over the gate voltage and magnetic field range. In the second pass, we fixed the value of the fit parameter $Q_c$ at $Q_{c,\mathrm{avg}}$, to reduce the spurious variations in the internal quality factor $Q_i$ yielded by the fitting algorithm. The end results as a function of field and gate are shown in Fig.\,\ref{supp_f0map_Qimap}.

\begin{figure}[htb]
\includegraphics{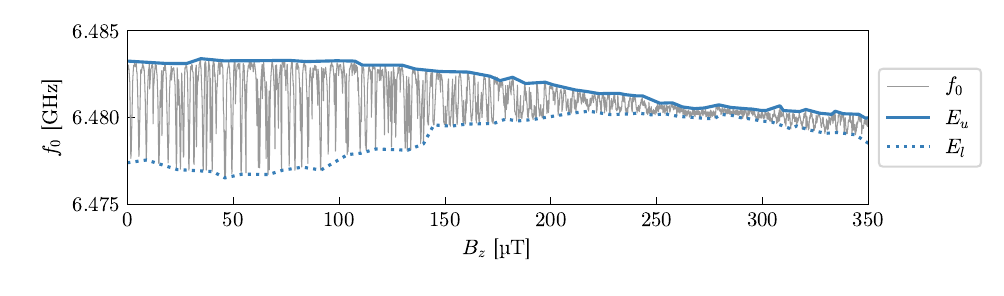}
\caption{\label{fig_envelopes} Resonant frequency ($f_0$) and envelopes ($E_u, E_l$) at $V_g=4\,$V (increasing B field).} Same data as in the inset of Fig.\,4(a), but in an extended range.
\end{figure}

The upper and lower envelopes of the $f_0(B_z)$ data points have been determined by searching for the maximum and minimum in a sliding window. We used a window size of 15 data points, which corresponds to $\Delta B_z=1.5\,$\textmu T. This window is large enough to contain the extrema of the fast oscillation, but not too large, so that the variation on the intermediate scale can be captured. 
Between the detected extrema, we interpolate linearly for visualization and calculation of the envelope size $\delta E$. In Fig.\,\ref{fig_envelopes} we show the determined envelopes as well as $f_0(B_z)$ in an extended range for the same data as in Fig.\,4(a) (blue curve).

\section{\label{sec:addata}Additional data}

\begin{figure}[htb]
\includegraphics{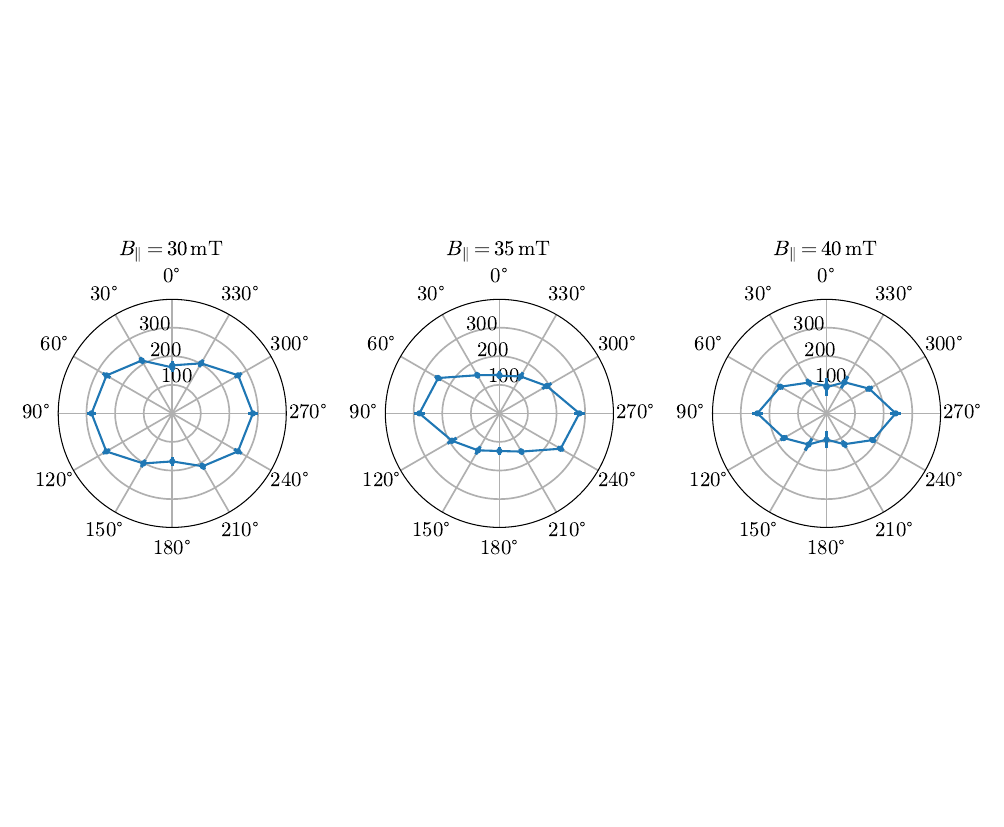}
\caption{\label{Ic_inplane} Critical current (in nA) as a function of $\theta$ (angle between $B_\|$ and the $x$-axis) in in-plane magnetic field for $B_\|=30, 35$ and 40\,mT. The radial line segments show the standard error.}
\end{figure}

In the main text in Fig.\,5(c) we show the suppression of the critical current along two directions of the in-plane magnetic field, the $x$ and $y$-axis. 
Fig.\,\ref{Ic_inplane} shows the angle dependence in the plane with a 30\textdegree~resolution.  When $B_\|$ is parallel to the direction of the supercurrent (and the rectangular contacts) the suppression is stronger\,\cite{Suominen2017}.

\bibliographystyle{apsrev4-2}
\bibliography{librarySM}